# The Internet Protocol

## Past, some current limitations and a glimpse of a possible future

Detailed summary of the Seminar
presented to the *Universidade da Beira Interior*
as partial requirement for the degree of Habilitation

*Sumário detalhado do Seminário*
*apresentado à Universidade da Beira Interior*
*como requisito parcial para admissão às provas de Agregação*

*by* Nuno Manuel Garcia dos Santos

Assinado por : **NUNO MANUEL GARCIA DOS SANTOS**
Num. de Identificação: BI069283346
Data: 2020.08.27 11:39:21 +0100

CARTÃO DE CIDADÃO

## *Sumário detalhado do seminário*

*O Protocolo Internet, o passado, algumas das suas limitações actuais e um vislumbre de possível futuro*

**The Internet Protocol, past, some current limitations and a glimpse of a possible future**

> *Apresentado em cumprimento do disposto no Art.º 8 alínea c) do Decreto-Lei 239/2007 de 19 de Junho, como requisito parcial para a admissão às Provas de Agregação em Engenharia Informática, na Universidade da Beira Interior, pelo candidato Nuno Manuel Garcia dos Santos.*

Nuno Manuel Garcia dos Santos

*Professor Auxiliar*

*Departamento de Informática*
*Faculdade de Engenharia*
*Universidade da Beira Interior*
*R. Marquês d'Ávila e Bolama*
*6201-001 Covilhã*
*Portugal*

# Abstract

The network layer is central to the networking scientific area. It is around the network layer that all the data communications develop, and one of its main tasks is to allow the identification of each single interface/machine between the potentially many interfaces in a network. This seminar addresses some of the issues that are usually presented to young Computer Science Engineering students in the course of several classes, but also presents some topics that are not address in networking courses. It is mostly focused on using Internet Protocol addresses in Local Area Networks, also considering issues that belong to the Wide Area Networks, such as data aggregation.

This document summarizes the content of a seminar, therefore it comprehends both teaching and researching subject. The seminar starts with a history of the evolution of the communication protocols from the early days of networks up until IPv6. It describes a new approach to define the addresses of network interfaces using Variable Length Subnet Masks, as usually this is a not an easy task for Computer Science Engineering undergraduate students.

This summary also describes some of the limitations of the data communication in todays' networks, proposing some solutions, where possible, including a novel mean of connectionless data transmission by using IPv6 addresses, by extension of previously published research.

The way the seminar is organized provides a history to the past of the Internet Protocol, a view of some of its well-known current limitations, and a glimpse into a possible future regarding an improved connectionless layer 3 data transfer protocol.

*(this page is intentionally left blank)*

# Table of contents

# Table Index

# Figure Index

# Abbreviations and acronyms

μs – microsecond
AFRINIC – African Network Information Centre
APNIC – Asia Pacific Network Information Centre
ARIN – American Registry for Internet Number
CIDR – Classless Interdomain Routing
CRC – Cyclic Redundancy Code
DAL – Destination Address Length
dK-IPv6 – Destination Keyed User Datagram Protocol
E-mail – Electronic mail
FCS – Frame Check Sequence
FTP – File Transfer Protocol
GB – Giga Byte
HTML – Hyper Text Mark-up Language
HTTP – Hyper Text Transfer Protocol
IAB – Internet Activities Board, *currently* Internet Architecture Board
IANA – Internet Assigned Numbers Authority
IBM – International Business Machines
IEFT – Internet Engineering Task Force
IMP – Interface Message Processor
IoT – Internet of Things
IP – Internet Protocol
IPng – Internet Protocol next generation
IPv *x* – Internet Protocol version *x*
KB – Kilo Byte
KUDP – Keyed Used Datagram Protocol
LACNIC – Latin America and Caribbean Network Information Centre
MB – Mega Byte
ms – millisecond
MPLS – Multi Protocol Label Switching
MTU – Maximum Transmission Unit
NCP – Network Control Protocol
netnews – Network News
NAT-PT – Network Address Translation – Port Translation
OSI – Open Systems Interconnect Model
PAC – Packet Aggregator and Converter
PBX – Private Branch Exchange *or* Public Branch Exchange
PC – Personal Computer
RFC – Request for Comments
RIPE NCC – Réseaux IP Européens Network Coordination Centre
RIR – Regional Internet Registries
SAL – Source Address Length
SDH – Synchronous Digital Hierarchy
SDN – Software Defined Networks
sdK-IPv6 – Source-Destination Keyed IPv6
sK-IPv6 – Source Keyed IPv6
sKUDP – Source Keyed UDP
SONET – Synchronous Optical Network
TCP – Transmission Control Protocol
TCP/IP – Transmission Control Protocol over Internet Protocol
UDP – User Datagram Protocol
VoIP – Voice over IP

(this page is intentionally left blank)

# 1. Introduction

This report is presented by Nuno Manuel Garcia dos Santos as a partial requirement for the public exams for the Degree of Habilitation as required by the Decree Law 239/2007 published in the *Diário da República, 1ª Série, Nº 116, de 19 de Junho de 2007*, in particular described in Article 5 c) and point 2. c) of Article 8.

This Detailed Summary of the Seminar addresses the subject of "The Internet Protocol, past, some current limitations and a glimpse of a possible future". The motivation for the selection of these topics is twofold: first, an effort was made to build educational content on the history of the development of the Internet Protocol, for reasons described forward; second, this Seminar will address some of the research lead by the candidate since 2004, namely, in the context of the Research Team of Siemens / Nokia Siemens Networks, of the European COST Actions IC0703 Traffic Measurement and Analysis, IC1303 Architectures, Algorithms and Protocols for Enhanced Living Environments, and also research done at the University of Beira Interior and at the *Instituto de Telecomunicações*.

Directly connected with this seminar, the contributions of the candidate for the last years can be summarized as follows:

- A patent on "Method and machine for aggregating a plurality of data packets into a unified transport data packet," International Patent WO 2007/118594 (A1), 2007; and European Patent EP 1848172 (A1), 2007;
- Publications reporting research on
  - the impact of the Ethernet frame payload size on IPv4 and IPv6 traffic;
  - the assessment of the efficiency of burst assembly / traffic aggregation algorithms, with real traffic data traces;
  - the new almost-reliable KUDP protocol and its efficiency tests;
- A new approach to the assignment of addresses to subnetworks using a Variable Length Subnet Masks.

The detailed summary is organized as follows[1]:

- the first chapter contains this short introduction;
- Chapter two addresses the historical perspective of the development of the IP protocols, including the two major versions of the IP protocol, version 4 and version 6;
- Chapter three presents the need to use classless IPv4 addressing and compares the current best practices with a novel method to assign addresses to IPv4 subnetworks. It also discusses the case of subnetwork addressing with IPv6;
- Chapter four presents some of the well-known limitation of the data transmission using the TCP/IP protocol stack;
- Chapter five presents the new Keyed IPv6 protocol that extends the Keyed UDP protocol;
- Chapter six presents conclusions and future work.

[1] *Expressions that are not in English appear italicized, and when appropriate, its translation to English is provided. Abbreviations and acronyms are provided in page iv.*

# 2. From telephone numbers to IP Addressing

## 2.1. Introduction

It is not uncommon that curricular units on Computer Networks to be so densely packed with content, that there is little room to discuss the history of the technology, and with that, teach the students who were / are the pioneers that have created / create the technological marvels that support a whole new manner for humans to live in society. This is not the main goal of this seminar, as we will not present all the names of the scientists and researchers that created the many protocols and standards that make the world of Computer Networks, but we will also not evade this issue and therefore, we will try to present the evolution of the addressing at the network layer in a historical perspective.

Addressing machines was something that happened naturally as a result of the expansion of the telephone network. Initially small PBX (Private Branch Exchange or Public Branch Exchange) facilities were run by human telephone operators who knew which extension belonged to whom and their work consisted in physically making the circuit that allowed the communication between the two telephones. These eventually grew larger and larger because more and more persons wanted to have a telephone device in their businesses and homes, making it impossible for the operator to know which socket connected to who's device in the telephone network. A website reports that the first telephone number was proposed in the middle of an outbreak of measles [1], when a worried physician suggested that his telephone should be associated with a number to allow his patients to easily memorize the terminal to which they should ask to be connected. When PBX started to interconnect with each other, the next step seems to have been to add an area prefix to the telephone number [2] as a mean to better identify to which region a given subscriber belongs. This would be the example that Internet addresses would follow, particularly because at that time both computers and telephones communicated using dedicated circuits.

Current addressing schemes, both IPv4 and IPv6 (references provided forward), are given to the students without a perspective of why these became to be how we currently know them. Yet, while teaching Computer Networks, one of the questions one hears the most is "Why are we working with IPv4? What happened to IPv1, 2, 3, and 5?".

Trying to teach IPv4 and IPv6 has taken such a large effort and time in classes in a typical undergraduate course, that the history of the development of the IP Protocol, and of course, the history of what happened to IPv0, IPv1, IPv2, IPv3 and IPv5 is very often never mentioned, nor are the names of those who pioneered this area.

This chapter provides an overview on the rationale of IP address concepts, with some detail on the why's, who's and when's of the journey from the Network Control Protocol in the early days of the Arpanet to the post-1996 Internet Protocol version 6.

### 2.2. The first IP Addresses are not version 1

In the early days of networking, the TCP (Transmission Control Protocol) had the spotlight regarding the deployment of networks. TCP had succeeded to NCP (Network Control Protocol) which had been deployed by the Network Working Group lead by Steve Crocker, and had seen its first release in December 1970 as the Host-to-Host protocol of ARPANET [3]. At that time, NCP did not have the ability to address networks nor machines, but it would rather send data to an IMP (Interface Message Processor), a computer that is better described as a very simple router. NCP relied on the physical structure of ARPANET to make sure the data packets were delivered to their destination, and when this did not happen, the applications would usually crash [3]. To solve these issues, Robert Kahn proposed to develop a new protocol that would eventually become parts of the TCP/IP protocol stack ("TCP/IP" meaning "TCP over IP), serving this more like a communication protocol and NCP would serve more like a device driver.

In the Spring of 1973 Kahn invited Vinton Cerf, who had been involved in the development of NCP, to join the team that would develop TCP, and in later that year, in a conference in Sussex, United Kingdom, a report with the basic concepts of TCP and later published as a paper [4], was distributed to the participants.

It is interesting to understand that, at this point, TCP was not so much oriented to deal with addressing, but rather to solving the packet loss problem, by implementing a control mechanism that involved windowed sends and acknowledgments. While focusing on control, TCP considered already a 4-byte address, being the first byte the network address and the remaining three bytes, the address of the host. At the time, having 256 possible networks was seen as more than enough. Yet, Ethernet was already under development at Xerox's Palo Alto Research

Center (PARC), and this technology would completely change the landscape regarding the number of networks [3] that would come to exist. The Transmission Control Protocol would eventually be published in 1981 in a RFC (Request for Comments) edited by Jon Postel [5].

Regarding versions, a memorandum from the Internet Assigned Numbers Authority (IANA) acknowledges that versions 0 to 3 of the Internet Protocol existed as experimental versions [6]. This memorandum also mentions the possibility of the existence of IPv7 up until IPv15, but this falls out of the scope of this Seminar, and as far as the author knows, there have been no further developments on versions higher than 6.

It is commonly assumed that IPv0 refers to the first TCP implementation and while IPv0 and IPv1 are referenced in this RFC [7], IPv2 is only mentioned in the Internet Protocol Specification draft (front page and page 10) [8]. It is interesting to see that IPv2 already considered a 4-byte Destination Address but only a 3-byte Source Address, and to observe that these fields appeared in this order in the map for the IPv2 header, something that would change in later versions. Yet, the length of those addresses was expected to be variable and to be recorded in two previous fields named DAL and SAL for "Destination Address Length" and "Source Address Length", respectively. It is clear from this text that the "(…) `first octet of an address [is] to be interpreted as a network identifier, and that the rest of the address identifies the host within that network` (…)." This Draft also stated that "(…) `The 8 bit network number, which is the first octet of the variable length address, has a value specified in RFC 739` (…). `In any case, the latest information can be obtained from Jon Postel.` (…)" (*sic*). As a curiosity, RFC 739 registers only 16 networks [8]. With the exception of [6], references to IPv3 could not be found elsewhere.

Eventually TCP evolved to become the TCP/IP protocol stack, and the reason for this seems to have been a comment Postel published as an Internet Draft [9] in 1977. Postel writes:

> "(…) `We are screwing up in our design of internet protocols by violating the principle of layering. Specifically we are trying to use TCP to do two things: serve as a host level end to end protocol, and to serve as an internet packaging and routing protocol. These two things should be provided in a layered and modular way. I suggest that a new distinct internetwork protocol is needed, and that TCP be used strictly as a host level end to end protocol. I also believe that if TCP is used only` in this cleaner way it can be simplified somewhat. A third item must be specified as well -- the interface between the internet host to host protocol and the internet hop by hop protocol. (…)".

It can be noted that in 1977, Postel writes "internet" with a lower case "i", while in RFC 791 (1981) Internet appears both with in upper-case and lower-case. The question on when was it decided that Internet should be written with a upper or lower case "I" falls also out of the scope of this Seminar, but a very interesting 2015 article featuring this discussion, published by Wired magazine, can be found here https://www.wired.com/2015/10/should-you-be-capitalizing-the-word-internet/ [10].

When TCP was separated from IP, the Internet Protocol version 4 was expected to supply all the addresses network computing would ever need, as an address space as large as $2^{32}$ addresses seemed to be more than enough to support all the networking needs of the community. As a mean of comparison, this would allow the assignment of an IP address to each living person on Earth in 1985, a time when the IBM PC had just made its debut in the American businesses and households.

IPv4 is first described in 1981 in RFC 791, edited by Postel [11]. Nevertheless, RFC 791 does not consider the use of what is known as Network Mask, instead, this RFC applies only what is currently called Classful addressing. The following excerpt from RFC 791 describes this procedure:

"(…) Addresses are fixed length of four octets (32 bits). An address begins with a network number, followed by local address (called the "rest" field). There are three formats or classes of internet addresses: in class a, the high order bit is zero, the next 7 bits are the network, and the last 24 bits are the local address; in class b, the high order two bits are one-zero, the next 14 bits are the network and the last 16 bits are the local address; in class c, the high order three bits are one-one-zero, the next 21 bits are the network and the last 8 bits are the local address. (…)".

Before moving on to the next point in this seminar, it must be noted that IPv5 is considered to be registered as the Internet Stream Protocol (ST, 1979) [12], which later evolved to the Experimental Internet Stream Protocol [13] (1990). It has been referred that IPv5 or ST was faced with the inevitability of IPv4 address exhaustion in a (then) near future, yet despite all the work done in this protocol, and later reused in the foundations of Voice over IP (VoIP), it was decided to discontinue its development [14].

RFC 791 mentions the first three classes of IPv4 address space, namely, Classes A, B and C, stating what defines each address. Table 1 shows the definition for the five classes of IPv4 address space, being Class D reserved for Multicast transmissions, and Class E to be considered as reserved for future use [15].

Some authors argue that Class A address range start at `1.0.0.0` and ends with `127.255.255.255`. The point is sustained with the argument that addresses whose first byte is zero refer to "a host in this network", *i.e.*, "address `0.x.y.z` refers to host `x.y.z` in this network". Nevertheless, from the author's point of view there are three arguments to support the classification of addresses in this range as Class A type addresses: *1.* RFC 870 on Assigned Numbers also assumes that these addresses are class A [16], the same happening with the RFCs that obsoleted RFC 870; *2.* addresses `127.x.x.x` and `255.255.255.255` are also reserved and yet are generally considered to be Class A (the first) and Class E (the last), and finally, *3.* it will be very strange to have a Classful definition of the IPv4 address space and leave out a slice of that address space as un-classified.

*Table 1 – Classful IPv4 address space definition.*

| Class | Initial Address (binary and decimal notation) | Final Address (binary and decimal notation) |
|---|---|---|
| A | **0**0000000.00000000.00000000.00000000<br>0.0.0.0 | **0**1111111.11111111.11111111.11111111<br>127.255.255.255 |
| B | **10**000000.00000000.00000000.00000000<br>128.0.0.0 | **10**111111.11111111.11111111.11111111<br>191.255.255.255 |
| C | **110**00000.00000000.00000000.00000000<br>192.0.0.0 | **110**11111.11111111.11111111.11111111<br>223.255.255.255 |
| D | **1110**0000.00000000.00000000.00000000<br>224.0.0.0 | **1110**1111.11111111.11111111.11111111<br>239.255.255.255 |
| E | **1111**0000.00000000.00000000.00000000<br>240.0.0.0 | **1111**1111.11111111.11111111.11111111<br>255.255.255.255 |

### 2.1. IPv4 private and public addresses

RFC 1918 [17] is the last RFC to mention private IPv4 address definition. It is interesting to observe that in 1996, authors of RFC 1918 considered that there could be 3 categories of IPv4 addresses. The first one would be used by the hosts that, inside a company network, would not need connectivity to the exterior. Their addresses would be unambiguous inside that network, but possibly ambiguous between networks in other companies. The second category would be used by the hosts that would require "(…) `access to a limited set of outside services (e.g., E-mail, FTP, netnews, remote login) which can be handled by mediating gateways (e.g., application layer gateways)` (…)" [17], and these hosts would still use company-wise unambiguous addresses, but probably ambiguous between companies. The third category would be used by hosts that need global network access layer, and these would use globally unique IP addresses. Addresses for the first and second categories are termed as private, and addresses for the third category are termed as public. So for hosts in the second category, the connectivity would be provided by means of applications at the top layer, which would in turn, connect to the exterior hosts and their services.

The first definition of this classification for Private and Public IPv4 addresses appears first in 1994 in the Informational RFC [18], that states:

"(…)
```
The Internet Assigned Numbers Authority (IANA) has reserved the
following three blocks of the IP address space for private networks:

     10.0.0.0        -   10.255.255.255
     172.16.0.0      -   172.31.255.255
     192.168.0.0     -   192.168.255.255

We will refer to the first block as "24-bit block", the second as
"20-bit block, and to the third as "16-bit" block.  Note that the
first block is nothing but a single class A network number, while the
second block is a set of 16 contiguous class B network numbers, and
third block is a set of 255 contiguous class C network numbers.
```
(…)".

To be noted that although Informational RFC 1597 describes the third block as "(…) `a set of 255 contiguous class C network numbers` (…)" (*sic*), in fact the third block contains 256 contiguous class C networks, from network `192.168.0.0` to network `192.168.255.0`.

Yet, the reason for the definition of public and private addresses has its deep roots in one fact and one consequence [19]: the fact was the exponential way the Internet grew in the early 90's, and the consequence was of course, the exhaustion of the IPv4 address space.

The definition of private addresses is not done without some controversy. Authors of Informational RFC 1627 [20] argue that authors in RFC 1597 did not follow the usual public review and approval either from the IETF or by the IAB (which at the time meant Internet Activities Board). The point that the use of networks 10 is considered harmful, follows a vision that is better stated in the words in the authors, as follows:

> "(…) A common -- if not universal -- ideal for the future of IP is for every system to be globally accessible, given the proper security mechanisms. Whether such systems comprise toasters, light switches, utility power poles, field medical equipment, or the classic examples of "computers", our current model of assignment is to ensure that they can interoperate.
>
> In order for such a model to work there must exist a globally unique addressing system. (…)" [20].

Advocating for a unique IP address space and for the global accessibility of every system that had an IP address, followed by the vision of connected and accessible toasters and light switches in the Internet would eventually become a possibility with IPv6 (at that time termed IPng for Internet Protocol next generation) and with the concept of the Internet of Things (IoT). Authors of RFC 1627 conclude that the apparent relief brought by RFC 1597 recommendations would bring complacency from a large part of the community that did not take the long-term view of the problem into consideration, and consequently, would cause a much faster IPv4 address exhaustion at some later date.

With the proposal for IPv4 private addresses set, the remaining addresses would be, of course, public. Yet, there are addresses that are reserved, and therefore, cannot be classified neither as public nor as private. This is the case of Class E addresses, the broadcast address `255.255.255.255`, the Local Loopback range, and of course, the range of addresses that have its first 8 bits as zero. Therefore, in the opinion of the author, the complete classful breakdown for the classification of IPv4 addresses is shown in Table 2.

*Table 2 – Classful IPv4 address space definition, including Reserved, Public and Private blocks.*

| Class | Initial Address (binary and decimal notation) | Final Address (binary and decimal notation) |
|---|---|---|
| A Reserved | 00000000.00000000.00000000.00000000<br>0.0.0.0 | 00000000.11111111.11111111.11111111<br>0.255.255.255 |
| A Public | 00000001.00000000.00000000.00000000<br>1.0.0.0 | 01111110.11111111.11111111.11111111<br>126.255.255.255 |
| A Private | 00001010.00000000.00000000.00000000<br>10.0.0.0 | 00001010.11111111.11111111.11111111<br>10.255.255.255 |
| A Public | 00001011.00000000.00000000.00000000<br>11.0.0.0 | 01111110.11111111.11111111.11111111<br>126.255.255.255 |
| A Local Loopback | 01111111.00000000.00000000.00000000<br>127.0.0.0 | 01111111.11111111.11111111.11111111<br>127.255.255.255 |
| B Public | 10000000.00000000.00000000.00000000<br>128.0.0.0 | 10101100.00001111.11111111.11111111<br>172.15.255.255 |
| B Private | 10101100.00010000.00000000.00000000<br>172.16.0.0 | 10101100.00011111.11111111.11111111<br>172.31.255.255 |
| B Public | 10101100.00100000.00000000.00000000<br>172.32.0.0 | 10111111.11111111.11111111.11111111<br>191.255.255.255 |
| C Public | 11000000.00000000.00000000.00000000<br>192.0.0.0 | 11011111.11111111.11111111.11111111<br>223.255.255.255 |
| C Private | 11000000.10101000.00000000.00000000<br>192.168.0.0 | 11000000.10101000.11111111.11111111<br>192.168.255.255 |
| C Public | 11000000.10101001.00000000.00000000<br>192.163.0.0 | 11011111.11111111.11111111.11111111<br>223.255.255.255 |
| D Multicast | 11100000.00000000.00000000.00000000<br>224.0.0.0 | 11101111.11111111.11111111.11111111<br>239.255.255.255 |
| E Reserved | 11110000.00000000.00000000.00000000<br>240.0.0.0 | 11111111.11111111.11111111.11111111<br>255.255.255.255 |

### 2.1. IPv4 notation for identification of the network and host addresses

According to RFC 1918, the slicing of private IPv4 addresses (then termed "Private Internets") was done respecting its original class, so for example, the first block of private addresses stands as single network with $2^{24}$ available addresses, the second block represents 32 networks of 65536 available addresses each, and the third block represents 256 networks each with 256 available addresses.

Regarding the management of the overall IPv4 address space, it seems that the attempt to solve two different problems resulted in what was to become CIDR, the initials standing for Classless Inter Domain Routing, an Address Assignment and Aggregation Strategy, with its first proposed standard being published in 1993 [21]. Yet, the first report of these problems was published in 1992, in the Informational RFC 1338, "Supernetting: an Address Assignment and Aggregation Strategy" [22]. Yu *et al.* described the problem as follows:

"(…) As the Internet has evolved and grown over in recent years, it has become painfully evident that it is soon to face several serious scaling problems. These include:

1. Exhaustion of the class-B network address space. One fundamental cause of this problem is the lack of a network class of a size which is appropriate for mid-sized organization; class-C, with a maximum of 254 host addresses, is too small while class-B, which allows up to 65534 addresses, is to large to be widely allocated.

2. Growth of routing tables in Internet routers beyond the ability of current software (and people) to effectively manage.

3. Eventual exhaustion of the 32-bit IP address space. (…)" [22].

It seemed clear that an address aggregation strategy would improve the problem of the exponential growth of the size of routing tables, this solution only disturbed by two situations, being these caused by companies that, as they switched from one Internet connectivity provider to another, would make "a hole" in the provider's address space, and by multi-homed companies, which by their nature require that all their addresses are advertised in the network.

Hierarchical routing also needed a more precise manner of defining which part of the address was the network part and which part was the host part. RFC 1338 propose the following change in the specification of the network and the host addresses [22]:

"(…) To simplify administration, it would be

```
useful if filter elements automatically allowed more specific network
numbers and masks to pass in filter elements given for a more general
mask.  Thus, filter elements which looked like:

     accept 128.32.0.0
     accept 128.120.0.0
     accept 134.139.0.0
     accept 36.0.0.0

would look something like:

     accept 128.32.0.0 255.255.0.0
     accept 128.120.0.0 255.255.0.0
     accept 134.139.0.0 255.255.0.0
     deny 36.2.0.0 255.255.0.0
     accept 36.0.0.0 255.0.0.0

This is merely making explicit the network mask which was implied by
the class-A/B/C classification of network numbers.
```

(…)" [22].

The introduction of the concept of "network + mask" pair is one of the innovations of this RFC, proposing that would be necessary to deploy new Inter-Domain routing algorithms capable of handling arbitrary "network + mask" pairs. This RFC also proposed that route `0.0.0.0` with mask `0.0.0.0` is to be used as default route and that this must be accepted by all implementations. The mask has been used since to identify what is the part of the address that corresponds to the network address, and what part corresponds to the address for the hosts.

The usually called "slash notation" did not appear for a while and in fact, in 1992, both RFC 1338 and RFC 1519 still referenced the address `192.0.0.0/8` as "`192.0.0.0/255.0.0.0`". Although being used in the industry for many years, the first RFC that uses the slash notation for IPv4 is RFC 4632 [23], published in 2006, reporting the best practice for Internet address assignment and aggregation plan.

In either case, the innovation of identifying the two zones of the address (network and host), allowed the implementation of strategies that were more efficient in using the IP address space. The operation that is performed on an address to determine what part of the address is the network address is known as "ANDing" and it is a simple logical AND operation between the IP address and the mask. Table 3 shows an example of the calculation of the network address, and by opposition, the host address of the network + mask pair `193.136.66.69 255.255.255.240`.

*Table 3 – Example for the calculation of a Network Address.*

| | | | | | | | |
|---|---|---|---|---|---|---|---|
| Address | 11000001<br>193 | . | 10001000<br>136 | . | 01000010<br>66 | . | 01000101<br>69 |
| Mask | 11111111<br>255 | . | 11111111<br>255 | . | 11111111<br>255 | . | 11110000<br>240 |
| ***AND*** | | | | | | | |
| resulting Network Address | 11000001<br>193 | . | 10001000<br>136 | . | 01000010<br>66 | . | 01000000<br>64 |

As RFC 4632 is dated August 2006, it seems logical to assume that the slash notation was first used in IPv6. In fact, although the IPv6 standard proposal was first published in December 1995, in RFC 1883 [24], the description of the proposed manner to represent an IPv6 address and its prefix was published in July 1998 [25]. It is interesting to observe the text in the original RFC, stating

> "(…) `The text representation of IPv6 address prefixes is similar to the way IPv4 addresses prefixes are written in CIDR notation. An IPv6 address prefix is represented by the notation:`
>
> `ipv6-address/prefix-length` (…)" [25].

## 2.2. IPv6 Address space

The version 6 of the Internet Protocol [24], [26] was designed with several concerns in mind, the more prominent being to define a protocol that would not suffer from address space shortage for a long time. In fact, the IPv6 header considers addresses that are 128 bits long, the first IPv6 address being `0000:0000:0000:0000:0000:0000:0000:0000`, also written as `::`, and the last address being `FFFF:FFFF:FFFF:FFFF:FFFF:FFFF:FFFF:FFFF`, allowing for a number of addresses that is too big to allow a common human to have a worldly notion of how big it is. But a comparison can be provided: currently there are some 7.700.000.000 ($7.7 \times 10^9$) humans living on Earth [27]; each human body has an estimated 3.7x10.000.000.000.000 ($3.7 \times 10^{13}$) cells [28]. As the number of IPv6 addresses is $2^{128}$, we could assign more than 1.2 million addresses to each and every cell for each and every human, but not just only for Earth, but also for the one thousand million worlds equal to ours.

The very large number of IP addresses is so big that it really loses meaning to many of us. This is perhaps why subnetting in IPv6 makes little sense, and we proposed that IPv6 should be listed as one of the first resources in the history of Engineering that is not scarse. Nevertheless, not being scarse does not mean that it does not have to be managed, and in fact it is. One of the first comments on IPv6 address management is the way the address space is organized, as not all the addresses are available for general use, as sub-section 2.3 details.

IPv6 considers that a part of the address refers to the network, and the remaining part refers to the interface address. This is a slight change in the way IPv4 addresses were conceptualized as the initial idea was "one address for one host", "one host" here standing for "one machine". Of course, for IPv4 there are hosts that have more than one IPv4 address, typically one address per network interface, *e.g.* an IPv4 router with several interfaces. For IPv6, the concept has changed to "one address per interface", but again, one network interface has more than one IPv6 address.

### 2.3. IPv6 notation for identification of the network and the interface addresses

IPv6 addresses use the slash notation to identify which part of the address is to be considered as the network address and which part addresses the interface in the network node. It has to be noted that the non-network part of the address is not expected to identify the node in the network, but rather the interface in a node in the network, therefore assuming that a particular node may have several different interfaces with different IP addresses. Also noteworthy is the fact that most operating systems now allow for an interface to have multiple IP addresses [29].

Therefore, an address such as `FE80:0:0:0:0:0:0:2/10` reserves the first 10 bits to the network address and the remaining 122 bits to the interface address. Yet, some specific types of IPv6 network addresses may have more complex architectures, and such an example are the Local IPv6 Unicast Addresses as described by RFC 4193 [30]. In this case, the `FC00::/7` addresses, described forward, are mapped as

| | | | | |
|---|---|---|---|---|
| 7 bits | 1 | 40 bits | 16 bits | 64 bits |
| Prefix | L | Global ID | Subnet ID | Interface ID |

where the Prefix allows the identification of the type of address, the flag `L` is set to `1` if the prefix is locally assigned (being set to `0` is reserved for future use), the `Global ID` is used to create a globally unique prefix, the `Subnet ID` is used to identify the subnetwork within the site, and the `Interface ID` is used to identify a given interface in a particular host in the network.

### 2.4. Text representation of IPv6 addresses

One of the issues regarding IPv6 addresses is that often a large number of zeros needs to be written. RFC 3513 [31] provides the rules to simplify the text representation of IPv6 addresses. In its section on "Text Representation of the Addresses", it is clearly stated that "(…) `it is not necessary to write the leading zeros in an individual field` (…)", and that "(…) `The use of "::" indicates one or more groups of 16 bits of zeros. The "::" can only appear once in an address. The "::" can also be used to compress leading or trailing zeros in an address.` (…)".

Therefore, the following addresses "(…)

```
      1080:0:0:0:8:800:200C:417A  a unicast address
      FF01:0:0:0:0:0:0:101        a multicast address
      0:0:0:0:0:0:0:1             the loopback address
      0:0:0:0:0:0:0:0             the unspecified addresses

   may be represented as:

      1080::8:800:200C:417A       a unicast address
      FF01::101                   a multicast address
      ::1                         the loopback address
      ::                          the unspecified addresses
```

(…)" (quoting directly from RFC 3513 [31]).

Yet, there can be the case where addresses have more than one representation. For example,

| | |
|---|---|
| the address | `12AB:0000:0000:CD30:0000:0000:0000:0000` |
| can be represented as | `12AB::CD30:0:0:0:0` |
| or as | `12AB:0:0:CD30::` |
| but ***not*** as | *`12AB::CD30::,`* |

as this would cause a violation of one of the rules defined by RFC 3513.

The representation of IPv4 addresses embedded in IPv6 addresses also has a particular form of notation that allows for a simpler identification of the address itself. RFC 3513 proposed two types of IPv4-in-IPv6 addresses, the “IPv4-compatible IPv6 Address” and the “IPv4-mapped IPv6 Address”. Although these are not very common, the first type would be used by hosts and routers to dynamically tunnel IPv6 packets over IPv4 routing infrastructure. RFC 3513 states that IPv6 nodes that use this technique are assigned special IPv6 unicast addresses that carry a global IPv4 address. The second type would be used to represent IPv4 addresses as IPv6 addresses.

The mapping for these addresses is as follows:

IPv4-compatible IPv6 Address

| 80 bits | 16 bits | 32 bits |
|---|---|---|
| 0000 … 0000 | 0000 | IPv4 Address |

IPv4-mapped IPv6 Address

| 80 bits | 16 bits | 32 bits |
|---|---|---|
| 0000 … 0000 | FFFF | IPv4 Address |

Their text representation can be done as these examples show:

“(…)

```
    0:0:0:0:0:0:13.1.68.3

    0:0:0:0:0:FFFF:129.144.52.38

 or in compressed form:

    ::13.1.68.3

    ::FFFF:129.144.52.38
```

(…)” (quoting directly from RFC 4291 [29]).

### 2.5. IPv6 types of addresses

Regarding the way the addresses are classified, it has to be noted that there are other changes in the way IPv6 communicates, when compared to IPv4. For IPv4, the communication is either unicast, multicast or broadcast, meaning the first, a communication from one address to another address, the second, a communication from one address to another address that is present in several machines (a Class D address), and the final, a communication from one machine to all the machines in that network, as routers do not route broadcast packets [21], [23].

Similar to IPv4, IPv6 can send unicast packets, and packets in multicast mode. Additionally, IPv6 provides a new communication manner, in which anycast packets are sent and are received by the first machine within a group of machines that are configured with the same address [26]. Unlike IPv4, not having the need for a broadcast address nor for a network address within a network address range, the IPv6 addresses can be used from the first value to the last value, *i.e.,* from the address where the bits in the interface zone are all zeros, to the address where the bits in the interface zone are all ones.

Combining the information in RFC 2373 and RFC 4193, IPv6 addresses are classified and grouped as follows [25][30]:

- Unspecified address `::/128`
- Loopback `::1/128`
- Unique local unicast `FC80::/7`
- Link-local unicast `FE80::/10`
- Multicast `FF00::/8`
- Global Unicast all other addresses.

Additionally, Informational RFC 3849 [32] proposes that the range of addresses `2001:db8::/32` should be marked by global routers as non-routable addresses to allow these to be used as documentation addresses. Authors of RFC 3849 claim that using this range of addresses reduces the probability of conflicts and confusion when IPv6 addresses are used in documentation, exercises or educational materials. This range of addresses was first accepted within the Asia Pacific Network Information Centre (APNIC), one of the five worldwide Regional Internet Registries (RIR) along with AFRINIC (Africa), ARIN (North America), LACNIC (Latin America) and RIPE NCC (Europe), and is now widely accepted.

Also noteworthy if the fact that RFC 3513 had proposed an additional type of local unicast address range termed “Site-local unicast”, that would be have the prefix `FEC0::/10`. Yet this range of addresses was to be deprecated 17 months later by RFC 3879 [33], with the argument that site-local addresses sometimes leaked over to global routers and caused applications to fail, among other arguments related to developer pain and manager pain while defining, using and maintaining site-local addresses.

However, as the introduction of RFC 3879 states, the decision to deprecate this type of addresses was not consensual ("(…) `the consensus was far from unanimous` (…)" [33]), and 13 months later, RFC 4193 [30] proposes an additional type of addresses termed Unique Local IPv6 Unicast Addresses. The abstract of this RFC states "(…) `This document defines an IPv6 unicast address format that is globally unique and is intended for local communications, usually inside of a site. These addresses are not expected to be routable on the global Internet.` (…)" [30]. Curiously enough, the authors of RFC 4193 state the idea of Unique-local addresses was not exclusive to them and credit the work of Brian Carpenter and Christian Huitema, who were the authors of RFC 3879.

Among the specificities of these types of local addresses is the use of additional information to the IPv6 address, representing the identification of the site and / or link to which it refers to. This is referred to as IPv6 address scoping and is defined by RFC 4007 [34]. Examples of addresses with scope information are:

"(…)

```
          fe80::1234 (on the 1st link of the node)
          ff02::5678 (on the 5th link of the node)
          ff08::9abc (on the 10th organization of the node)

   would be represented as follows:

          fe80::1234%1
          ff02::5678%5
          ff08::9abc%10

   (Here we assume a natural translation from a zone index to the
   <zone_id> part, where the Nth zone of any scope is translated into
   "N".)

   If we use interface names as <zone_id>, those addresses could also be
   represented as follows:

         fe80::1234%ne0
         ff02::5678%pvc1.3
         ff08::9abc%interface10
```

(…)" (directly quoting [34]).

The zone delimiter identifier is useful when, for example, a user needs to connect via a specific link within a particular host that has several links with the same IPv6 local address. Continuing the example, if a router has several interfaces with the address `fe80::1`, connecting to the host at the other end of the *e.g.* 2nd interface in that router, would require the application handling this request to process specific information. Therefore, the destination address would be `fe80::1%2`.

Following the architecture of IPv6 addresses described in [25], there are some changes regarding the structure of IPv4 addresses, namely,

i) in IPv4 there was a range of addresses for Local Loopback, the `127.0.0.0/8`, and for IPv6, there is only one Loopback address, the `::1/128`;
ii) the concept of Address Classes was dropped;
iii) and finally, there are no group of addresses that are reserved for future use, as there was for IPv4 addresses in Class E.

### 2.6. Functions of each group of addresses in IPv6

Three of the four groups of IPv6 addresses have similar functions to groups of addresses in IPv4. In IPv4 addresses were separated in 5 classes, A to E, and in each of the A, B and C classes, there were private addresses and public addresses, being the private addresses non-routable outside their respective networks, and the public addresses the ones that are globally routable.

IPv6 addresses in the ranges `FC80::/7` and `FE80::/10` are not expected to be globally routable and therefore can be said to be private addresses. Yet of these two groups, only the addresses in the range `FC80::/7` are considered to be equivalent to private IPv4 addresses, as the `FE80::/10` are private in a sense that is more strict, as these are not expected to exist outside a layer two broadcast domain, *i.e.*, these addresses exist only in a given link.

Addresses in the range `FF00::/8` are used as multicast addresses, having a direct correspondence in terms of functionalities with addresses in Class D for IPv4.

With the exception of the previous address ranges and of the unspecified and loopback addresses (`::/128` and `::1/128`), and also considering the range of documentation addresses (`2001:db8::/32`), all other IPv6 addresses are public and globally routable.

### 2.7. Conclusions

This chapter was dedicated to providing a historical overview of the evolution of the addressing needs in networks, starting with the natural heritage of the addresses in the telephone network, describing the two major Internet Protocol addressing schemes, IP version 4 and IP version 6.

The goal of this chapter was multiple. Primarily, the reader can understand the details on IP addressing for each of the major versions, identifying the functionalities of an address based on its format and content. Also a goal of this chapter is to provide a historical perspective on the evolution of the addressing in the network. Often in Computer Science the names of the pioneers remain unacknowledged, and in the opinion of the author of this Seminar, due credit has to be made.

One of the issues that was not referenced in this chapter was the issue of IPv4 public address exhaustion. Also not mentioned, but worthy of another seminar are the issues related to the coexistence of IPv4 and IPv6, and the efforts made by the global network engineering community towards the adoption of IPv6.

Despite these efforts, IPv4 still plays a major role in the network engineering field, and more so in what are usually called local networks. Finally, and because of this, this chapter serves as introduction as the 3$^{rd}$ chapter, where a technique to identify IPv4 addresses for multiple subnetworks in a local network is presented, discussed and compared with the current best practices.

# 3. Classless network addressing in IPv4 and the case for IPv6

## 3.1. Introduction

Although IPv6 was introduced in 1998 as a upgrade to IPv4, there was never an intention to replace IPv4, and at the technical *fora*, the expression used is "IPv4 to IPv6 transition" [35]–[37]. Currently, IPv4 is still widely used in local networks and throughout the Internet [38], and it is expectable that it continues to be so, as is it more convenient to use and teach, and easier to programme in network devices.

This chapter describes the use of IPv4 addresses in a classless manner, presents the current best practice and a novel method to assign network addresses to several subnetworks.

## 3.2. From classful to classless network addressing

As mentioned in Chapter 2, IPv4 was initially designed to be used in a classful manner. For example, Class B would allow 65536 networks each comprising 65534 hosts. It soon became clear that the manner the IP address space was being used was not very efficient, thus allowing for the introduction of the concept of classless addressing, in which an address is used independently of the class it belongs to.

Regarding the range of private addresses, as seen in Table 2, each of classes A, B and C have a range of private addresses, summarized here as follows:

Class A - `10.0.0.0 .. 10.255.255.255.255`

Class B - `172.16.0.0 .. 172.31.255.255`

Class C - `192.168.0.0 .. 192.168.255.255`.

Also worth noting is that in classful addressing, Classes A, B and C use network masks of 8, 16 and 24 bits respectively, *i.e.*, these classes use network masks represented as `255.0.0.0`, `255.255.0.0` and `255.255.255.0`.

It has to be noted that for any range of addresses in IPv4, the first and the last address of the range are reserved as the address of the network itself and the address for broadcast messages. Therefore, if a network has 16 bits available as host address space, the case of the private address range for Class C, the number of usable addresses *N* is

$$N = 2^n - 2, \quad (1)$$

where N is the number of usable addresses, and $n$ is the number of bits available as host space in the address.

The use of addresses that are flexible and do not need to comply with the separation between network and host address defined by their respective classes allows network managers to use the IPv4 scarse address space more efficiently.

Of course, it makes no sense to discuss classless nor classful subnetting in IPv6, as IPv6 does not classify its address space into classes. Yet, and similar to IPv4, it makes sense to discuss subnetting in IPv6, as both addresses are organized in network address + {host | interface} address.

### 3.3. Addressing Subnetworks

When a given network needs to be reorganized in several different subnetworks, the address space needs also to be rearranged. There are several reasons to define subnetworks inside a network, including to increase security, to increase the performance of the network by decreasing the size of the broadcast domains and to allow for better organization of the network, usually by assigning different subnetworks to different areas or departments inside the organization. Different subnetworks can still communicate with each other by forwarding their traffic to a router, which in turn can implement other rules for increased security and traffic flow optimization.

In large organizations it is common to have also a large number of subnetworks, and of course these need to be distinguishable between each other to allow the efficiency of routing tasks and the enforcement of security rules. Having different address spaces for different subnetworks also improves the ability of network managers to better manage network equipment and to detect and correct equipment malfunctions.

Consider the network in Figure 1. If the network administrator was to use only the private address space of Class A in a classful manner, the task could not be completed, as Class A private address space comprehends only one network with a maximum number of $2^{24} - 2$ hosts.

However, the network administrator could use four private network addresses from Class B, assigning for example the address space `172.16.0.0/16` to the network connected by `Switch0`, the address space `172.17.0.0/16` to the network connected by `Switch1`, and the address spaces `172.18.0.0/16` and `172.19.0.0/16` to the networks connected by `Switch2` and `Switch3`, respectively. In this case, each of these address spaces allows for 65534 addresses, and as each of the networks only needs to address 40 hosts (plus the address of the interface on the router), it is obvious that there is a large portion of the addresses that will remain unused.

Using Class C private network address spaces proves to be more efficient, as for example, the network administrator could use the addresses `192.168.0.0/24`, `192.168.1.0/24`, `192.168.2.0/24` and `192.168.3.0/24` to assign to each of the networks in each of the switches. Again, as each of these subnetworks allows for $2^8 - 2$ maximum number of hosts, *i.e.*, 254 hosts, and as the requirements are to address 41 hosts, there is still a large address space that remains unused.

So in summary, the whole network needed at the most 164 addresses and ended up using 262144 addresses if using classful addressing in Class B, and 1024 addresses if using classful addressing in Class C, resulting in efficiency ratios of 0.06% for Class B address space usage and 16.02% efficiency ratio for Class C address space usage.

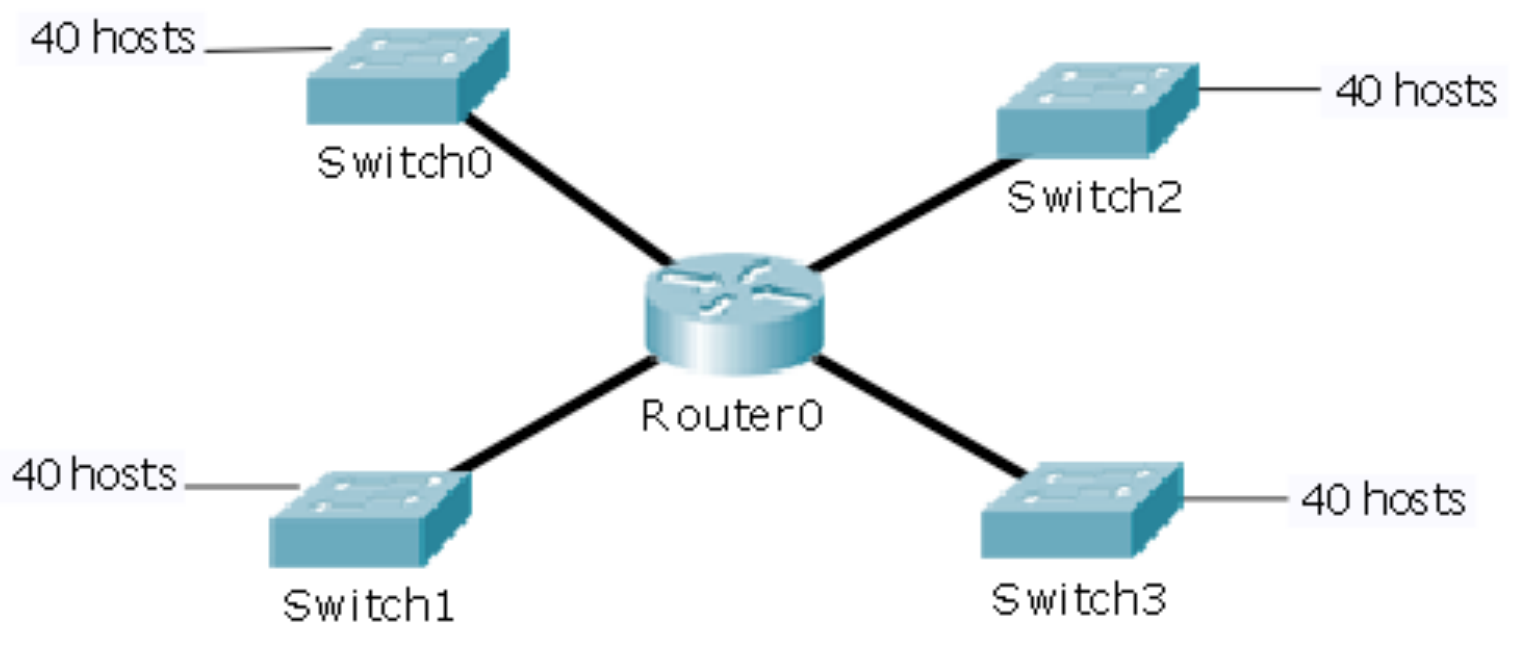

*Figure 1 – Sample network with 4 subnetworks with equal dimensions.*

Yet, if the network administrator decides to use classless addressing, it can be more efficient regarding the use of the address space. For the sake of example, let us say that the network administrator decides to use a Class C private address range, *e.g.*, `192.168.0.0`. As each of the networks needs to address 40 hosts plus the interface at the router, it needs 41 addresses. Using equation (1), it can be seen that to address 41 interfaces we will need 6 bits as $2^6 = 64$ and as $2^5 = 32$, so 5 bits are not enough but 6 bits allows the use of 64 addresses, a little more than the 41 + 2 addresses each network requires (41 addresses for the interfaces, plus the network and the broadcast address).

As the address chosen is a Class C, it can be said that network masks can be at the least a `/24`, being possible to use network masks that are `/25` or higher (Table 3 provides an example). In summary, as we have used a Class C address, we have 24 bits or more for the network address and we have 8 bits or less for the hosts addresses. As we only need 6 bits for addressing each of the hosts in each of the networks, we can write that the networks connected by each of the switches can be addresses as `192.168.0.0/26`, `192.168.0.64/26`, `192.168.0.128/26` and `192.168.0.192/26`. By using classless addressing, we have used 256 addresses when in fact we only needed 164, *i.e.*, we still have many unused addresses, but this is by far the most efficient manner to address the networks in this sample scenario, as the address space usage efficiency is 64.06%, much better that any of previously analysed classful scenarios.

The case discussed in the previous paragraph would also hold true whatever class of addresses was chosen as a starting point. For example, choosing addresses from Class A could result in the subnetworks being addressed as `10.0.0.0/26`, `10.0.0.64/26`, `10.0.0.128/26` and `10.0.0.192/26`, still using 256 addresses out of the address space.

### 3.4. The case for a classless network addressing scheme: a sample problem

There are several methods that a student can use to perform classless addressing for subnetworks. The most common is the one proposed by Cisco Networking Academy, resourcing to what is usually called the “Magic Number” [39]. Without expressly referring to it, that was the method used in the example of addressing the subnetworks in Figure 1.

After identifying the number of addresses necessary for satisfying the requirements of the networks (and do take into consideration that in these examples, all networks have similar addressing requirements), we concluded that we needed 6 bits to address the hosts and the router interface. Because we have four different networks, we will need to use 2 bits to address each of the networks, being the network connected to `Switch0` addressed as `00`, being `01`, `10` and `11` the addresses for the other networks connected to `Switch1`, `Switch2` and `Switch3`, respectively, although this order is completely arbitrary. Quite conveniently, the 2 bits necessary to address each of the networks plus the 6 bits of the host address space sum 8 bits, allowing us to use any class of addresses we wish. Let us stay with Class C and let us use the same arbitrary address we used previously `192.168.0.0/24`. We now have that, the scheme of the addresses for this network is

| | | | | | | | |
|---|---|---|---|---|---|---|---|
| Address | 11000000<br>192 | . | 10101000<br>168 | . | 00000000<br>0 | . | 00000000<br>0 |

and as previously calculated, the most suitable network mask is a /26, *i.e.*,

| | | | | | | | |
|---|---|---|---|---|---|---|---|
| Network mask | 11111111<br>255 | . | 11111111<br>255 | . | 11111111<br>255 | . | 11000000<br>192 |

being the "Magic Number" calculated as 256 - 192 = 64. The "Magic Number" can now be used to give addresses to each of the network address slices in the addressing scheme, being 0 the first slice, 0+64 the second slice, 0+64*2 the third slice and 0+64*3 the fourth and final slice, *i.e.*, `192.168.0.0/26`, `192.168.0.64/26`, `192.168.0.128/26` and `192.168.0.192/26`.

Different network sizes result in the calculation of different "Magic Numbers", as the examples in Table 4 show. Using larger subnetwork sizes still makes this procedure possible, but as it is unrealistic, it will not be discussed.

*Table 4 – Calculating the "Magic Number" for different sample network scenarios.*

| Bits in the network mask | Magic Number | Address sequence |
|---|---|---|
| 11111111.11111111.11111111.10000000 | 256 - 128 = 128 | 0, 128 |
| 11111111.11111111.11111111.11100000 | 256 - 224 = 32 | 0, 32, 64, 96, 128, 160, 192, 224 |
| 11111111.11111111.11111111.11110000 | 256 - 240 = 16 | 0, 16, 32, 48, 64, 80, 96, 112, 128, 144, 160, 176, 192, 208, 224, 240 |

### 3.5. The case for a Variable Length Subnet Masks for network addressing scheme: a more realistic sample problem

The case in Figure 1, where all the subnetworks have similar addressing requirements is clearly a particular case and a simplification of any real word scenario. Consider now the sample network depicted in Figure 2. The scenario now considers 4 subnetworks, each requiring 4, 98, 13, and 49 addresses respectively for the subnetworks connected by `Switch0`, `Switch1`, `Switch2` and `Switch3`, respectively.

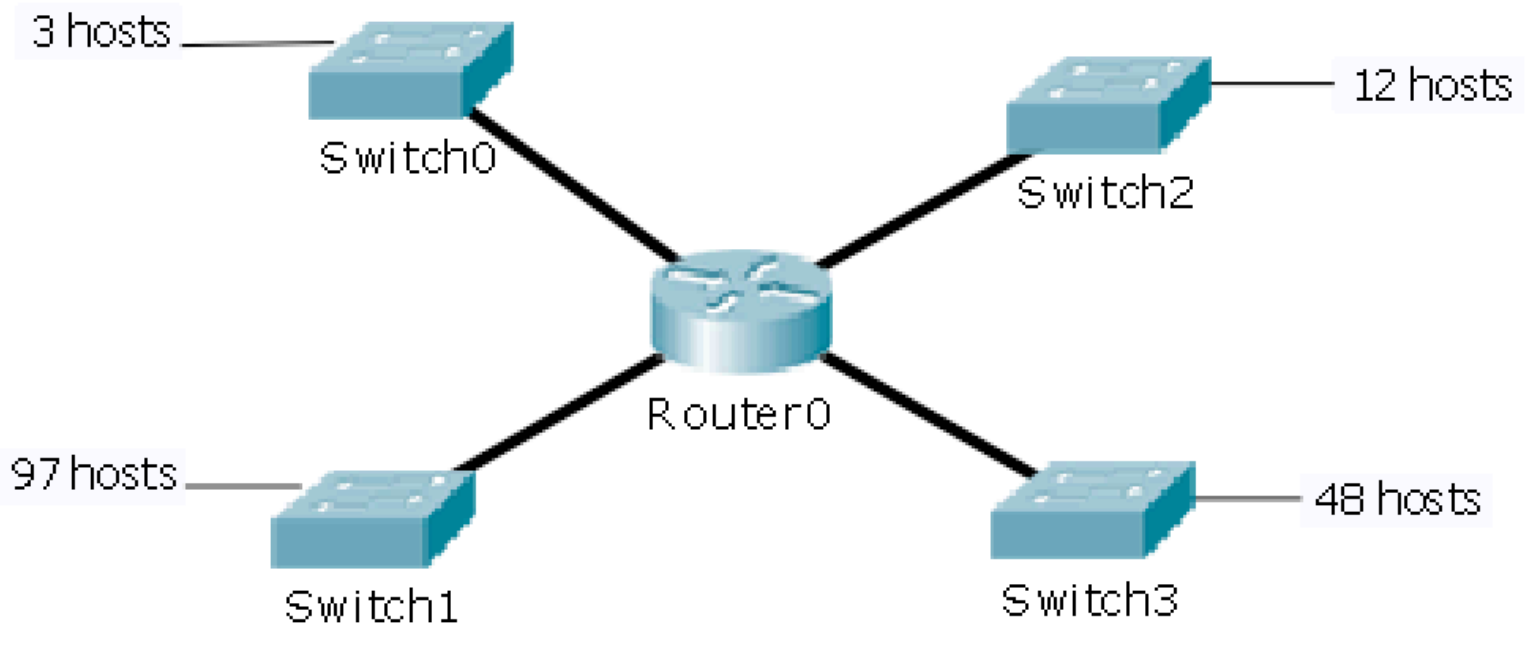

*Figure 2 – Sample network with 4 subnetworks with different dimensions.*

In a scenario where all subnetworks have different addressing requirements, the use of a "Magic Number" to identify the addressing schemes is inadequate as it would result in inefficiency. The use of the Variable Length Subnet Masks [40], [41] allows the definition of address ranges that minimize the waste of unused addresses. The procedure for this implies a number of steps, being the first step is to calculate the correct address space for each subnetwork. The calculations are shown in Table 5.

*Table 5 – Variable Length Subnet Mask for a sample network.*

| Subnetworks connected by | Required number of host addresses | Subnet mask | Awarded number of host addresses |
|---|---|---|---|
| `Switch1` | 98 | /25 | 128 |
| `Switch3` | 49 | /26 | 64 |
| `Switch2` | 13 | /28 | 16 |
| `Switch0` | 4 | /29 | 8 |

Arbitrarily considering the initial network address to be `192.168.0.0/24`, as used by previous example, we then have that the "Magic Numbers" for each of the subnetworks are 128, 192, and 208, resulting these from the sum of the initial address (`0`) with the number of awarded number of host addresses as seen in in Table 5. This results in the following addresses for each of the subnetworks: `192.168.0.0/25` for the network connect by `Switch1`, `192.168.0.128/26` for `Switch3`, `192.168.0.192/28` for `Switch2` and `192.168.0.208/29` for `Switch0`. The next available network range starts at `192.168.0.216`, but it can only comprehend additional networks that have `/29` masks, as other network masks would not produce valid addresses as we will show in the next section.

### 3.6. A new approach to Variable Length Subnet Mask

The case where the subnetworks required different address spaces is the norm rather than the exception, therefore, the use of Variable Length Subnet Masks (VLSM) is advisable to increase the efficiency of the use of the IP address space. This section presents a systematic approach to the use of VLSM that is novel and does not require the calculation of "Magic Numbers".

This method consists in the filling of a table that contains all the information required by a network administrator to assign addresses to subnetworks. This table is to be filled with the help of an auxiliary table, containing data related to the different possible network masks (Table 6). Table 6 does not list all the possible network mask lengths, starting at `/22` and ending at the largest possible network mask, `/30`.

*Table 6 – Network masks and number of possible host addresses.*

| Mask length | Number of available host addresses | Network Mask |
|---|---|---|
| … | … | … |
| `/22` | 1022 | `255.255.252.0` |
| `/23` | 510 | `255.255.254.0` |
| `/24` | 254 | `255.255.255.0` |
| `/25` | 126 | `255.255.255.128` |
| `/26` | 62 | `255.255.255.192` |
| `/27` | 30 | `255.255.255.224` |
| `/28` | 14 | `255.255.255.240` |
| `/29` | 6 | `255.255.255.248` |
| `/30` | 2 | `255.255.255.252` |

The number of available host addresses is calculated by using equation (1), where $n$ is calculated by subtracting the mask length from the size of an IPv4 address. For example, a /25 network mask allows only for 7 bits for the host address, thus the number of available addresses for the hosts is $2^7 - 2 = 126$.

The table with the addresses of the subnetworks has eight columns containing the necessary data to support the decision of the network administrator. The columns are described as follows:

1. Reference for the subnetwork (Ref)
2. Number of required host addresses (#Hosts)
3. Best suitable network mask (NM)
4. Number of awarded host addresses (from Table 6) (#AA)
5. Network address (NAddr)
6. Address of the first host (1st addr)
7. Address of the last host (last addr)
8. Broadcast Address (Bdcast).

In addition to Table 6, the network administrator also needs to know the initial available address and the size of the usable address space. The table is built in a line-by-line manner, using simple calculations and resorting to the data of the size of address space for each line. The first step while producing the table is to place the different subnetworks in each line, in descending order of the required number of hosts.

Each of the columns in the address table is calculated as follows:

1. **Ref** is the name / reference of sub network. The subnetworks appear on the table in descending order of size. This data is provided as part of the initial problem;
2. **#Hosts** is the number of required hosts. This data is also provided as part of the initial problem;
3. **NM** is the network mask that is the best fit to support the required number of hosts. For example, if the number of required hosts is 25, the best fit would be to use a network mask of `/27` as this network can address 30 hosts (data obtained from Table 6);
4. **#AA** is the number of awarded host addresses. This data is also obtained from Table 6;

5. **NAddr** is the network address. The network address in the first line is part of the problem given data. Subsequent network addresses (in subsequent lines) are obtained by adding `1` to the **Bdcast** address in the previous line;
6. **1st addr** is the address of the first host and it can be calculated by adding `1` to the network address;
7. **Last addr** is the address of the last host and it can be calculated by adding the value in **#AA** to the **NAddr**;
8. **Bdcast** is the broadcast address for this subnetwork and it can be calculated by adding `1` to the value of **Last addr**.

Table 7 shows an example of an empty Subnetwork Addressing Table. The cells marked in green background contain data that is provided by the text of the addressing problem, the cells marked with blue background are obtained by looking up data from Table 6, and the cells marked with orange background are obtained through simple calculations. Table 8 shows the same table, but here the operations are depicted by graphic elements. In a general manner, for the *i*th line, the first two cells contain values obtained from the problem statement, the 3rd and 4th cells contain data that was looked up in the network mask table (Table 6), the 5th cell contains the value obtained by adding `1` to the address in the 8th cell of the *(i-1)*th row, and the 6th, 7th and 8th cells contain addresses that are calculated as shown.

*Table 7 – Subnetwork addressing table using VLSM (empty table with calculations described).*

| Ref | #Hosts | NM | #AA | NAddr | 1st addr | Last addr | Bdcast |
|---|---|---|---|---|---|---|---|
| (given) | (given) | *in* Table 6 | *in* Table 6 | *(given address)* | NAddr+1 | NAddr+#AA | Last addr + 1 |
| (given) | (given) | *in* Table 6 | *in* Table 6 | Bdcast + 1 | NAddr+1 | NAddr+#AA | Last addr + 1 |

*Table 8 – Subnetwork addressing table using VLSM (empty table with calculations depicted).*

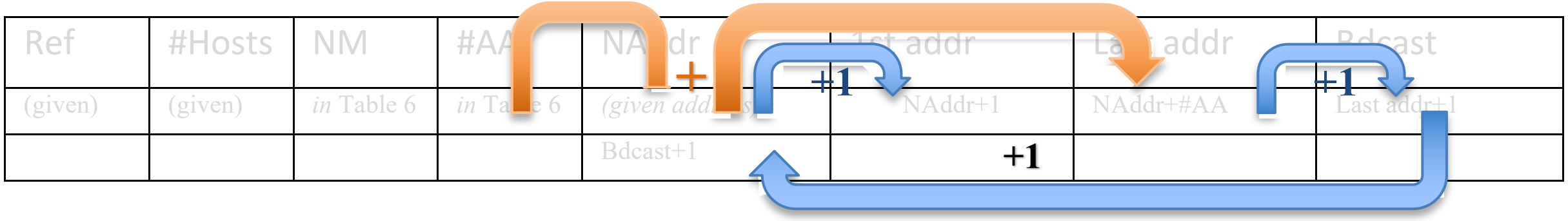

The subnetwork address table for the network shown in Figure 2, results as follows (Table 9). As it can be seen, all addresses in the Network Address and the Last Address columns are even, and all the addresses in the 1st address and the broadcast address are odd. Moreover, it can be demonstrated that all addresses comply with the rules for these specific addresses, being these: a) all network addresses have zeros in the host portion of the address, b) all 1st host addresses are 1, considering only the section of the host address, c) all the last addresses are composed by ones, except the Least Significant Bit that is zero, for the portion of the host address, and finally, d) all broadcast addresses are composed by ones regarding the portion of the address that is reserved for the hosts.

*Table 9 – Subnetwork addressing using VLSM (sample network).*

Base network address: `192.168.0.0/24`

| Ref | #Hosts | NM | #AA | NAddr | 1st addr | Last addr | Bdcast |
|---|---|---|---|---|---|---|---|
| Switch1 | 98 | /25 | 126 | 192.168.0.0 | 192.168.0.1 | 192.168.0.126 | 192.168.0.127 |
| Switch3 | 49 | /26 | 62 | 192.168.0.128 | 192.168.0.129 | 192.168.0.190 | 192.168.0.191 |
| Switch2 | 13 | /28 | 14 | 192.168.0.192 | 192.168.0.193 | 192.168.0.206 | 192.168.0.207 |
| Switch0 | 4 | /29 | 6 | 192.168.0.208 | 192.168.0.209 | 192.168.0.214 | 192.168.0.215 |

Let us consider a new example. In this example, we want to give addresses to *e.g.* six subnetworks, named from A to F with the following arbitrary address space requirements:

| Subnetwork | A | B | C | D | E | F |
|---|---|---|---|---|---|---|
| Required number of hosts | 10 | 2 | 50 | 80 | 22 | 30. |

The subnetworks can now be placed in the table in descending order of the number of required host addresses. Also arbitrarily, let us consider that we have been given the network address `10.0.0.0/24` to provide addresses to these six subnetworks. This value will be the network address of the first network and can also be added to the table at this point (Table 10).

*Table 10 – Subnetwork addressing using VLSM (another sample network).*

| Ref | #Hosts | NM | #AA | NAddr | 1$^{st}$ addr | Last addr | Bdcast |
|---|---|---|---|---|---|---|---|
| D | 80 | /25 | 126 | 10.0.0.0 | | | |
| C | 50 | /26 | 62 | | | | |
| F | 30 | /27 | 30 | | | | |
| E | 22 | /27 | 30 | | | | |
| A | 10 | /27 | 30 | | | | |
| B | 2 | /30 | 2 | | | | |

Before we proceed to fill the table in, we need to verify that the provided address space can fulfil the requirements of all the subnetworks. We see that the original base network address is a `/24`, therefore allowing a set of 256 addresses. Yet, adding all the values in the column of Awarded Addresses, we see that these add up to 280, larger than the address space of 256 addresses that was initially considered. With more rigour, to this value we should add 12, representing 6 subnetworks times 2 reserved addresses per subnetwork, to conclude that we would in fact need 292 addresses. We must therefore conclude that, stated as is, this problem has no solution. But if we increase the base network address to 10.0.0.0/23, allowing for 1024 addresses, we can now solve the addressing problem.

Using this new set of data, we can now fill the address table as follows (Table 11).

*Table 11 – Subnetwork addressing table using VLSM (another sample network using as base address 10.0.0.0/23).*

| Ref | #Hosts | NM | #AA | NAddr | 1st addr | Last addr | Bdcast |
|---|---|---|---|---|---|---|---|
| D | 80 | /25 | 126 | 10.0.0.0 | 10.0.0.1 | 10.0.0.126 | 10.0.0.127 |
| C | 50 | /26 | 62 | 10.0.0.128 | 10.0.0.129 | 10.0.0.190 | 10.0.0.191 |
| F | 30 | /27 | 30 | 10.0.0.192 | 10.0.0.193 | 10.0.0.222 | 10.0.0.223 |
| E | 22 | /27 | 30 | 10.0.0.224 | 10.0.0.225 | 10.0.0.254 | 10.0.0.255 |
| A | 10 | /27 | 30 | 10.0.1.0 | 10.0.1.1 | 10.0.1.30 | 10.0.1.31 |
| B | 2 | /30 | 2 | 10.0.1.32 | 10.0.1.33 | 10.0.1.34 | 10.0.1.35 |

This approach allows the solution of subnetwork addressing problems using VLSM in a logical and straightforward manner. Moreover, and in the opinion of the author, this approach is novel regarding the manners it presents the data and it facilitates the understanding of VLSM subnetworking tasks.

### 3.7. The case for IPv6

There was never an effort to apply VLSM to IPv6, as the abundance of addressing resources is so flagrant, that this issue is never mentioned in the literature. As an anecdote, while it is usual to IPv6 Internet Service Providers to allow the user to manage a /64 address, the whole IPv4 space (32 bits) fits $2^{32}$ times in each IPv6 address given to the user.

The application of VLSM for IPv4 carries some restrictions, in particular, it is expected that the amount of machines / interfaces to address is fairly well defined at the address planning stage, which is something that sometimes does not happen, either because of the Network Manager lack of experience or clairvoyance, or because of organizational changes. When the foreseen sub-network needs are exceeded, this causes the Network Administrator to redesign the whole addressing scheme, or to consider the new required addresses to form a new sub-network. In IPv6 the risk of running out of addresses is not null, but today it is clearly unrealistic.

As a result, when planning the IPv6 address scheme for sub-networks, the Network Administrator has a much higher stack of resources to use and can afford to not be strict, usually granting each network a generous portion of the address space.

### 3.8. Conclusions

This chapter described the issues regarding classless addressing in IPv4, with a special focus on the discussion of the techniques used to define addresses to subnetworks. The current best practice was presented and discussed, and a novel method of addressing this problem was presented. The case for IPv6 was also discussed.

The chapter presented several examples solved by the application of "Magic Numbers" and also by the use of a subnetwork addressing table, that as far as the author is aware, constitutes a new approach to this process.

# 4. Some well-known limitations of the data transmission using the IP/TCP protocol stack

## 4.1. Introduction

Being the result of an evolution, the IP protocol is usable in what became known as the TCP/IP (TCP over IP) protocol stack. Currently implemented in all major Operating Systems, and following the layered approach defined by Jon Postel in 1977, of the TCP/IP comprises 4 layers, that can be mapped to the seven layers of the conceptual Open Systems Interconnect Model (OSI) [42]. The four layers of the TCP/IP protocol stack are:

- Closer to the user, the user data is encoded (if necessary) and this data unit is said to belong to the Application layer. Data in this layer is later encapsulated inside a Segment;
- The data segment comprises the data from the user and adds a segment header that has one or two basic functions, depending on whether the connection is said to be Connection Oriented or Connectionless. For connectionless segments, this layer provides source and destination addresses that are referred to as Port Numbers (in Networks) or as Socket Numbers (in Operating Systems and Programming). For connection oriented segments, additionally to the source and destination port numbers, the header also includes other information that allows for a two-way sequenced communication to be established, therefore tagging each data unit with a sequence number that allows the receiving machine to detect when a segment has been received out-of-sequence or is missing, therefore allowing solving the problem, including reordering the segments of asking the source machine for the retransmission of the lost data. The source and destination port number addresses are also often referred to as application specific identifiers, which is not completely correct as a single application can use more than one single port number, *e.g.*, a HTTP communication often start multiple HTTP requests to the HTTP server, each one being identified by a distinctive port number. The segment is encapsulated inside an Internet Protocol packet;
- The Internet Protocol layer adds another header to the segment. This header, among other information includes the source and the destination IP addresses, allowing for the data to be sent to the destination interface across the network. Nevertheless, before being sent outside the source machine, the packet has yet to be encapsulated inside a frame;

- the Data Link Layer, in which data frames are transmitted from one interface to the other in the same local network, include, among other fields, the frame header, the Internet Protocol packet, here termed as frame payload, and a frame closing Frame Check Sequence, that allows the receiving machine to assess if the communication was performed with errors, in which case, the frame is discarded with no additional action being taken.

This chapter addresses some well-known problems related to the data transmission in the TCP/IP model. In all fairness, it has to be stated that because of the layered approach to data communication, none of the problems are related to the Internet Protocol layer in itself, but rather to the interactions between each of the layers.

### 4.2. The 1500 Maximum Transmission Unit

It is interesting to understand that a large section of the technology that supports our digital communications, and in particular, the TCP/IP protocol stack, has been defined more than 20, sometimes 30 or 40 years ago. For example, the Ethernet technology, later standardized by the IEEE as standard IEEE 802.3 (and others followed), was published in 1980, 40 years ago. Up to this day, the layout for the Ethernet frame that the vast majority of the computers use is exactly the same as the one defined in 1980.

This of course has its advantages and drawbacks. One of the major advantages is that the plethora of devices that use Ethernet rely on a well-established, well-researched, and to many accounts, successful technology. One of its major drawbacks is the extremely limited capacity of the Ethernet payload, supporting at the most, 1500 bytes.

In 1980, a 1500-byte payload capacity was more than enough, given the nature of the data at the time. Also, there was the need to occupy the transmission channel for as little time as possible, as early Ethernet networks relied on a shared medium to be used by all the terminals, and collisions did / do happen at the physical medium when two or more terminals try to transmit and their data happens to occupy, totally or partially, the data unit transmission slot. By keeping the transmission units small, the chances for a collision to occur in the medium were also small [43]. Another reason to keep the payload at 1500 bytes maximum, is that if this was to be larger, the efficiency of the error detection function provided by the Cyclic Redundancy Code (CRC) information transmitted in the Frame Check Sequence (FCS) would not be assured, therefore possibly resulting in an event where an error would go unnoticed [43][44].

The nature of data in digital communications has changed immensely in the last 20 years, let alone the last 40 years [45], and much of today data is related to some type of rich content, being this image, audio and/or video [46]. The point being, precisely, that the architecture of an Ethernet frame, was designed to support the type of traffic and the technologies that were known in 1980, *i.e.*, email and FTP, as HTTP, HTML and documents containing rich content would not be a reality until more than a decade later. So, supporting a payload of a minimum of 46 bytes and a maximum of 1500 bytes was clearly enough for the data transmission scenarios in the 80's.

But in the era of all-things-digital, recording, processing, storing and transmitting large amounts of data is the norm rather than the exception. In a random example, a simple photography taken with a commodity digital camera or a smartphone, can easily be 4.5 MB in size, even with good multimedia compressing algorithms, this meaning that uploading this photography to a remote server in the network, will generate, at the least, 3000 data packets. In reality, this scenario can actually be worse in the cases where the photography is sent via HTTP or via email, as these formats often require the binary data in the photography file to be converted to a compatible format before it can be transmitted.

Network traffic analysis was done in [38]. A total of 42 GB of data in 36 files randomly chosen from a 268 data trace set of files were used in this procedure, from four different collection sites at the United States:

- the America's Path (AmericasPath) exchange point at Miami, Florida (AMP files),
- the Front Range GigaPOP which is a consortium of Universities, non-profit corporations and government agencies that cooperate in an aggregation point called the FRGP in order to share Wide Area Networking (WAN) services (FRG files),
- the University of Memphis, which has a POS (Packet over SONET) logical OC-3c link to Abilene's KSCY at Kansas City (MEM files),
- and the Texas universities GigaPOP at Rice University, Houston, Texas (TXS files)).

Of the analysed 42 GB of publicly available data trace files, authors concluded that only 0.52‰ of the recorded packets were larger than 1500 bytes, and assumed that this was a very probable annotation or recording error. The relevant remark about this observation is that although IP packets (both *v4* and *v6*) have a 2 byte long field to register the IP packet length, therefore allowing for IP packets to be as large as 65355 bytes, the 99.948% of the IP packets travelling in the network are at the most 1500 bytes long, *i.e.*, 2.28% their maximum possible size.

From the application / user perspective, using a payload at 2.28% of its maximum capacity strikes as major inefficiency for the data transmissions tasks. Yet, from the physical layer perspective, the limitation seems unavoidable, as it is imposed by the technology available in 1980.

### 4.3. Repetitive routing tasks

As a consequence of the discussion in section 4.2, it results that a single action from a user can generate several thousand of packets, all of which will have the same source and destination IP addresses, and moreover, the same source and destination port addresses. In fact, one of the major tasks performed by the Transport layer (Layer 4 on the OSI Model), is to segment the data generated in the upper layers and segment it as to generate IP packets that can fit into a Data Link data unit (*e.g.* an Ethernet frame). This is also the reason why the protocol data unit for the Transport layer is termed a "segment".

Routing each and every one of the thousands of packets that go from one source machine to a destination machine seems also to be very inefficient, as this means multiple repeated routing operations, including routing table queries, packet header processing, packet buffering, and so on.

In core networks, other formats that allow for faster routing / switching are used, for example, IP/MPLS, SDH/SONET or Carrier Ethernet [47]. But as technologies from Layer 1, 2 and 2.5 are out of the scope of this seminar, we can refocus the discussion on routing efforts.

Burst switched networks were proposed in 1983 as a way to achieve "improved bandwidth efficiencies" [48]. In 2008, authors in [49] have proposed a packet aggregation and disaggregation technique that can be used in routers and core network equipment that eliminates the need to route a group of packets that is being transmitted from one core router to the another. The concept is straightforward and relies on the concept of data aggregation. Also, and similar to routers which can perform data fragmentation *e.g.* when a link has a smaller MTU than the data ingress link, the Packet Aggregator and Converter machines (PAC-machine for short) also need to know how to perform the packet aggregation and disaggregation at the ingress and egress nodes.

The process is described in detail in [50]: an ingress router handling a fairly amount of incoming traffic performs the data burst aggregation. This means that the router waits for a given period of time for incoming traffic, creating as many output packet-queues as packet destinations or routes. The queued packets are preceded by an IP header (*v4* or *v6*), that contains all the untouched packets, back to back, as payload. If using an IPv6 format, there is even the possibility of using an IPv6 Jumbogram [51], that can carry a payload of up to 4 GB in size.

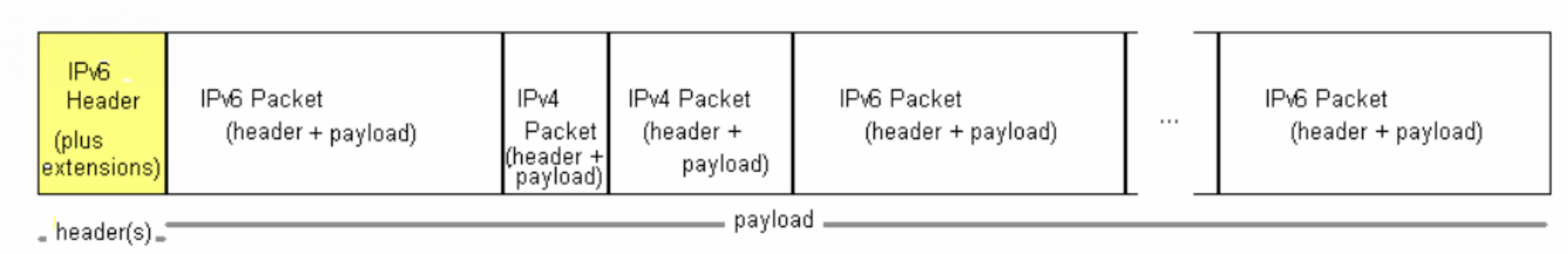

*Figure 3 – Aggregation of packets formed by an IP-PAC machine (from [50]).*

At the ingress node, an IP-PAC machine creates bursts using an hybrid burst assembly algorithm such as the one described in [52]. The IP-PAC machine would then proceed to send the bursts formed as seen in Figure 3 to the backbone / transport network. At the egress node, another IP-PAC machine performs the disaggregation of the packet burst, by simply interpreting and consequently discarding the leading IP header (Figure 3), and interpreting the payload bytes as a series of IP packets, eventually forwarding each of these packets to their respective exit interfaces towards their respective access networks. In the core network, the intermediate routers would only see a well-formed long IP packet.

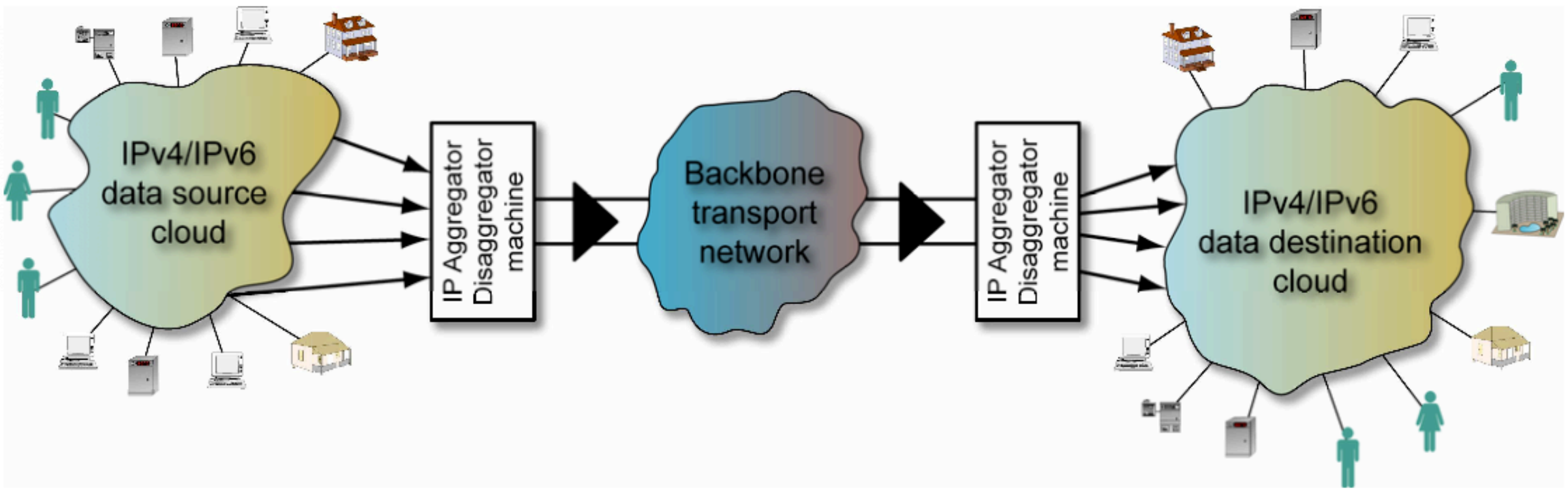

*Figure 4 – IP-PAC machines as gateways to and from a backbone network (from [50]).*

This technique has several advantages and several drawbacks. Among the drawbacks we have that buffering packets for the burst assembly and disassembling, increases the delay in the packets and interferes with the packet Jitter. Also, transmitting large packets may be an issue for the routers that do not have an adequate amount of buffering memory. Among the benefits of this technique we can count a reduced number of routing operations, and a decreased overhead because of the reduction of the waiting time between packets, and despite the addition of the PAC-machine packet header.

Regarding the process of assembling these packet bursts, research was done in [38][52][53] using real data traces, with burst-sized, time-limited, and both time and size constraint algorithms, these later also termed hybrid-burst-assembly algorithm. Using real network data, research concluded that keeping the maximum packet delay to an acceptable 10 ms, the aggregated packets can reach up to 9 KB in size, yet these values vary with the nature and load of the tributary data link. Research published in [52] also refers that for a time threshold of 10 ms, each burst is likely to have somewhere between 50 and 150 packets, depending on the considered data link traffic load. In these scenarios this would mean a significantly smaller number of routing tasks, between 50 to 150 times less routing tasks. Very likely, having a decrease in the number of routing tasks could result in having faster routing operations. Nevertheless, this trade-off was not subject of further research.

### 4.4. The impact of a larger IPv6 header

On a related research, and understanding that a significant portion of IP packets were in fact 1500 bytes in size, authors in [38] studied the effect that having a larger header on the IP packet would cause to the data traffic profile, as it would be the case for an IPv6 packet. The rationale for this research was as follows: if an IP packet is limited to a maximum of 1500 bytes because of the payload size of the Ethernet frame it will be transmitted in, having a larger header at the IP layer will decrease the usable space for the segment at the Transport layer. The underlying question was defined as “how significant is the impact of this increase in the overall network traffic profile”, specifically, “will this mean that there will be more packets flowing in the network?”.

Using the same data traffic traces collected from four different location in the United States, the authors of [38] proposed a packet construction reversion algorithm, in order to infer the characteristics of the original Application layer payload and consequently synthesize new IP packets with different header formats and constraints.

In this research, packets were grouped using source and destination IP addresses, source and destination Port numbers, and a time vicinity, *i.e.*, is a series of packets share the same pairs of source and destination IP and Port addresses and were transmitted very close in time, then probably, these packets were generated by the same action at the Application layer. Figure 5 and Figure 6 show a scheme depicting the payload reconstruction algorithm while manufacturing IPv6 packets.

Using the algorithm to infer the original application data size, new packets were then generated using different formats. Overall, almost 51 million packets were processed. The time vicinities for considering that a number of packets was generated by the same user-event were researched as 100 μs, 500 μs and 1 ms.

Additionally, two size thresholds were defined, trying to answer to the following questions:

a) “If the size limit at the source machine was defined to 9 KB instead of 1500 B, how many packets would that machine generate?”; and
b) “If the size limit at the source machine was defined to be 64 KB instead of 1500 B, how many packets would that machine generate?”.

These two size thresholds are, respectively, the maximum payload size for an Ethernet Jumbo Frame [54] and the maximum size of an IP packet [11], [24].

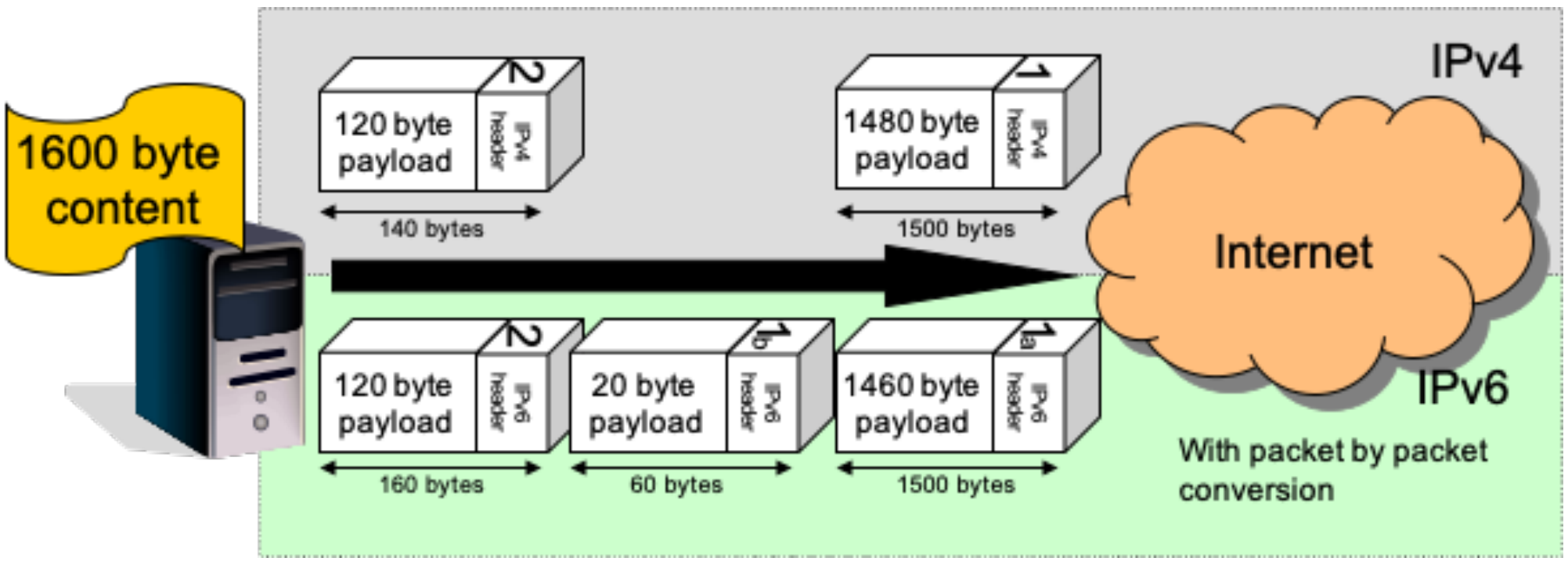

*Figure 5 –Packet remanufacturing algorithm performing a packet-by-packet conversion (from [38]).*

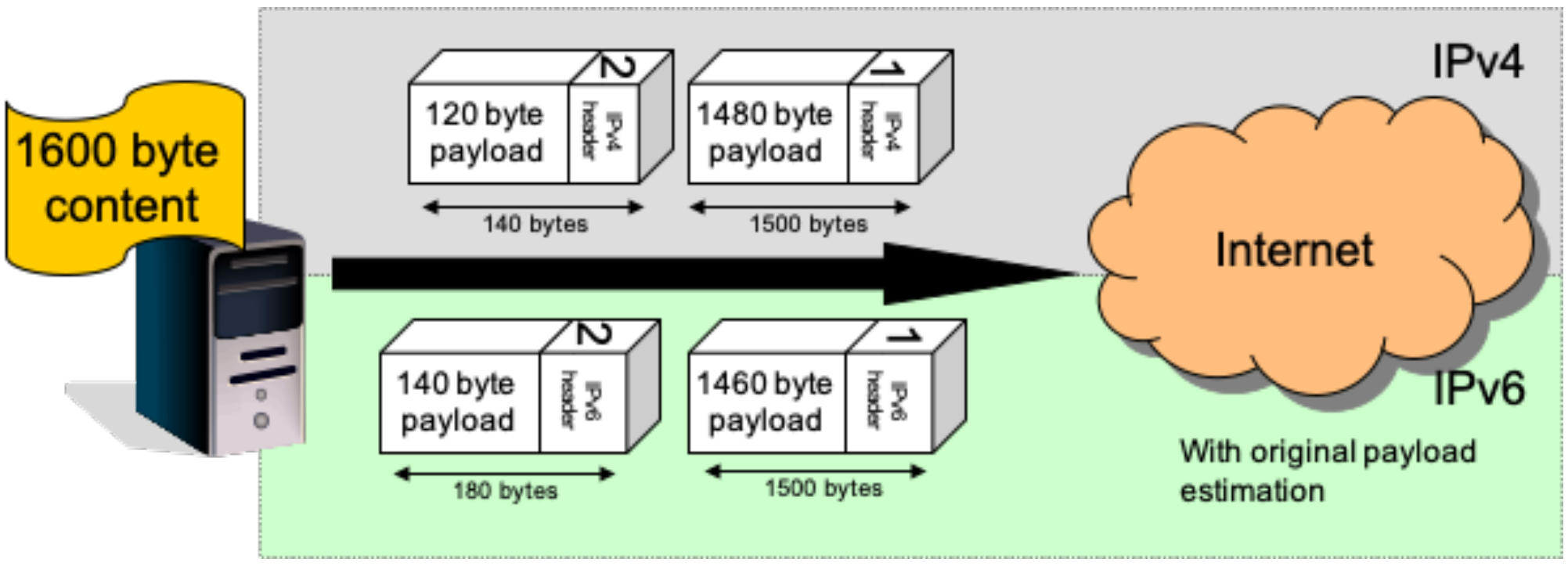

*Figure 6 –Payload reconstruction algorithm performing a conversion using the estimation of the original content (from [38]).*

Results in [38] report that a simple IPv4 to IPv6 header replacement could cause an increase in the number of packets flowing in the network that can range from 15% to 45%, depending on the data link load. Authors of [38] consider that this is a non-irrelevant increase and that it should be acknowledged as a draw-back resulting from the use of IPv6 instead of IPv4. Figure 7 shows a chart with the results for the header replacement algorithm depicted in Figure 5.

Additionally, and considering the unlikely changes in the size of the source-generated packets, [38] reports that for a time horizon of 500 μs the number of generated packets can be around 50% less if the packet size limit is defined as to comply to an Ethernet Jumbo Frame, and around 52%, as observed from combining Figure 8 and Figure 9. Moreover, if the size of the packet is defined to be at the maximum 64 KB and the time horizon is set to 1 ms, the number of generated packets at the source machine would be almost 60% less, as seen in Figure 10.

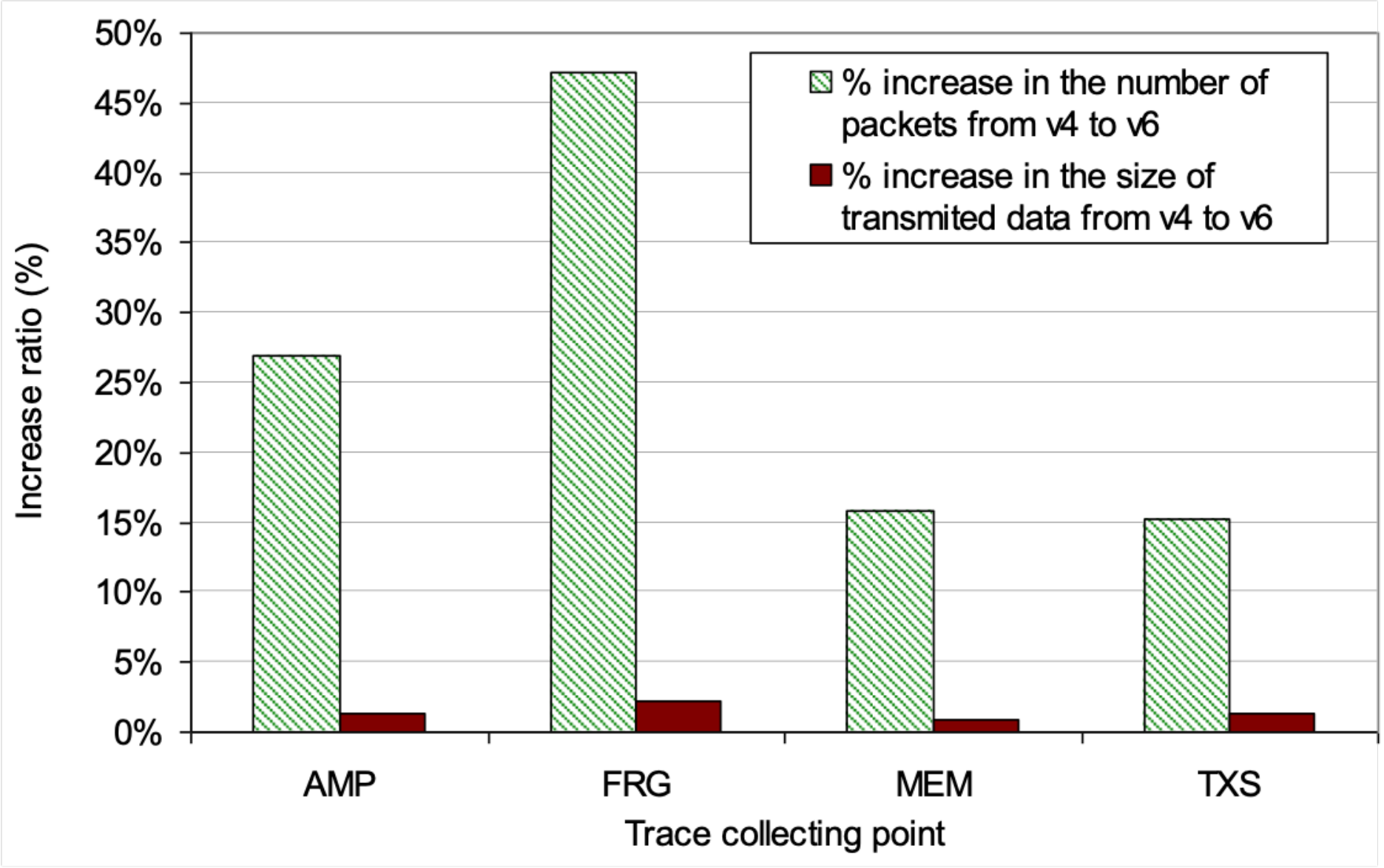

*Figure 7 – Increase on the number of packets and increase in the percentage of transmitted data resulting for using IPv6 and not IPv4 (from [38]).*

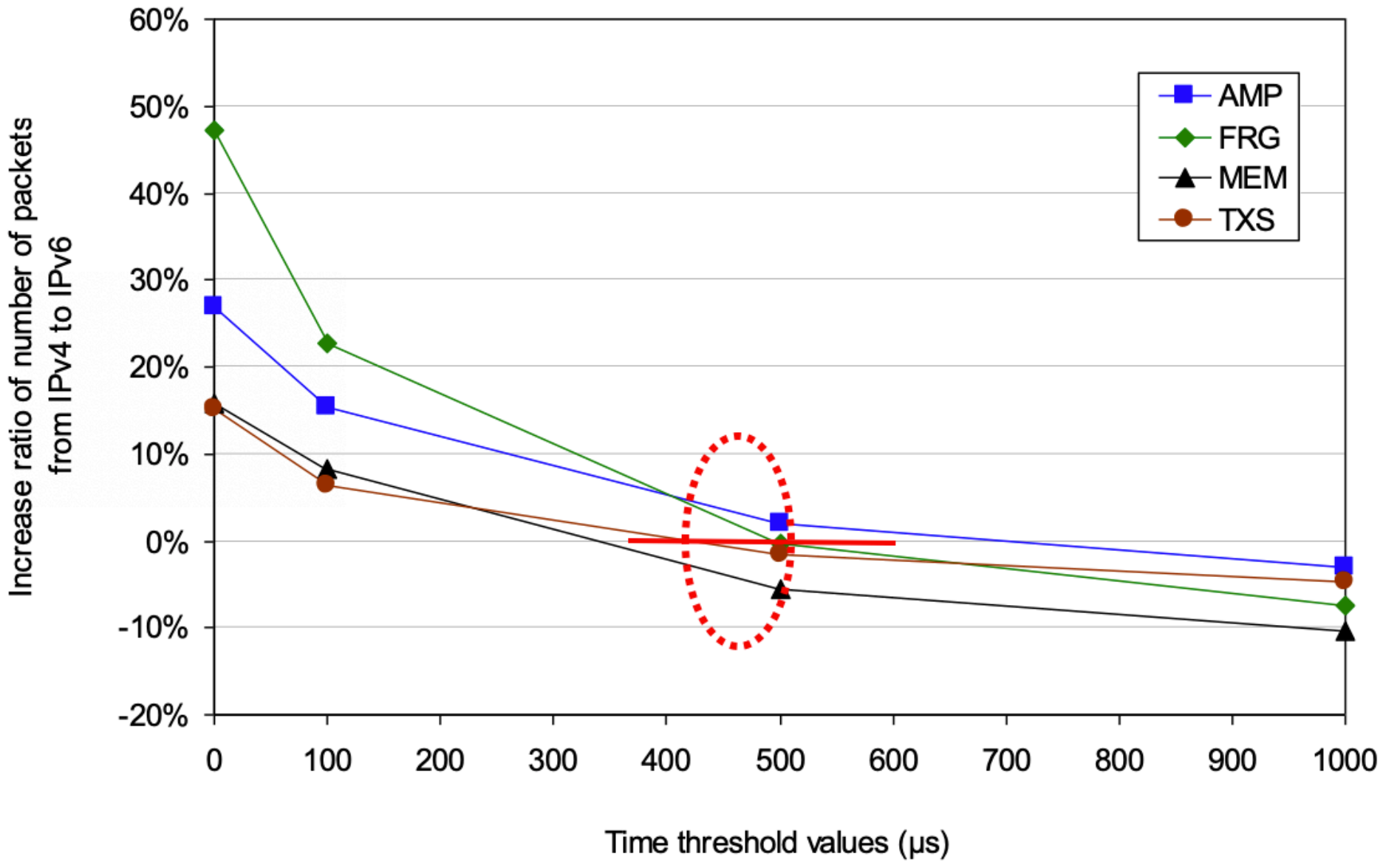

*Figure 8 – Different time aggregation thresholds and resulting increase / decrease in the number of generated packets, when converting IPv4 to IPv6, for a maximum packet size of 1500 B for each data trace (from [38]).*

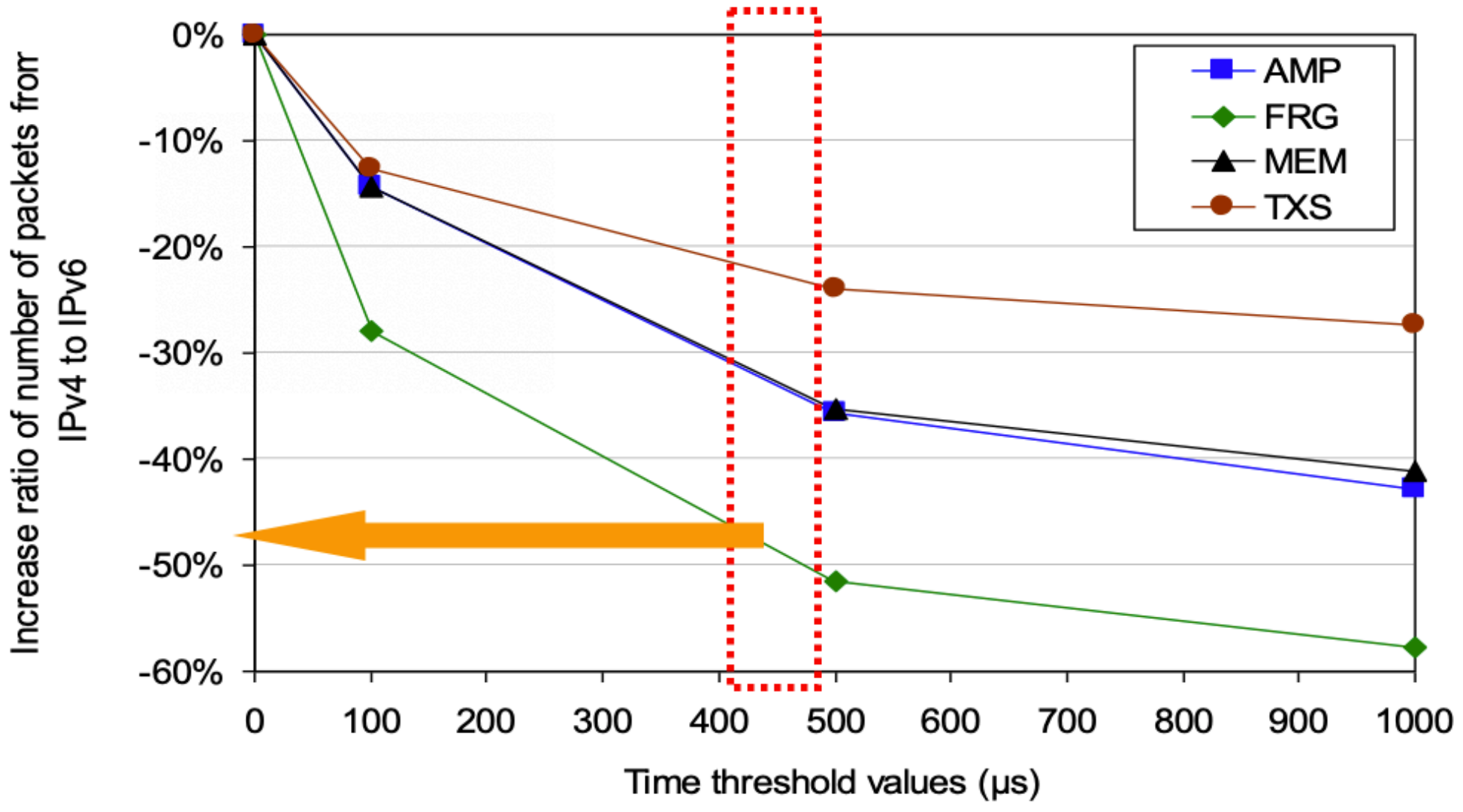

*Figure 9 – Different time aggregation thresholds and resulting increase / decrease in the number of generated packets, when converting IPv4 to IPv6, for a maximum packet size of 9 KB for each data trace (from [38]).*

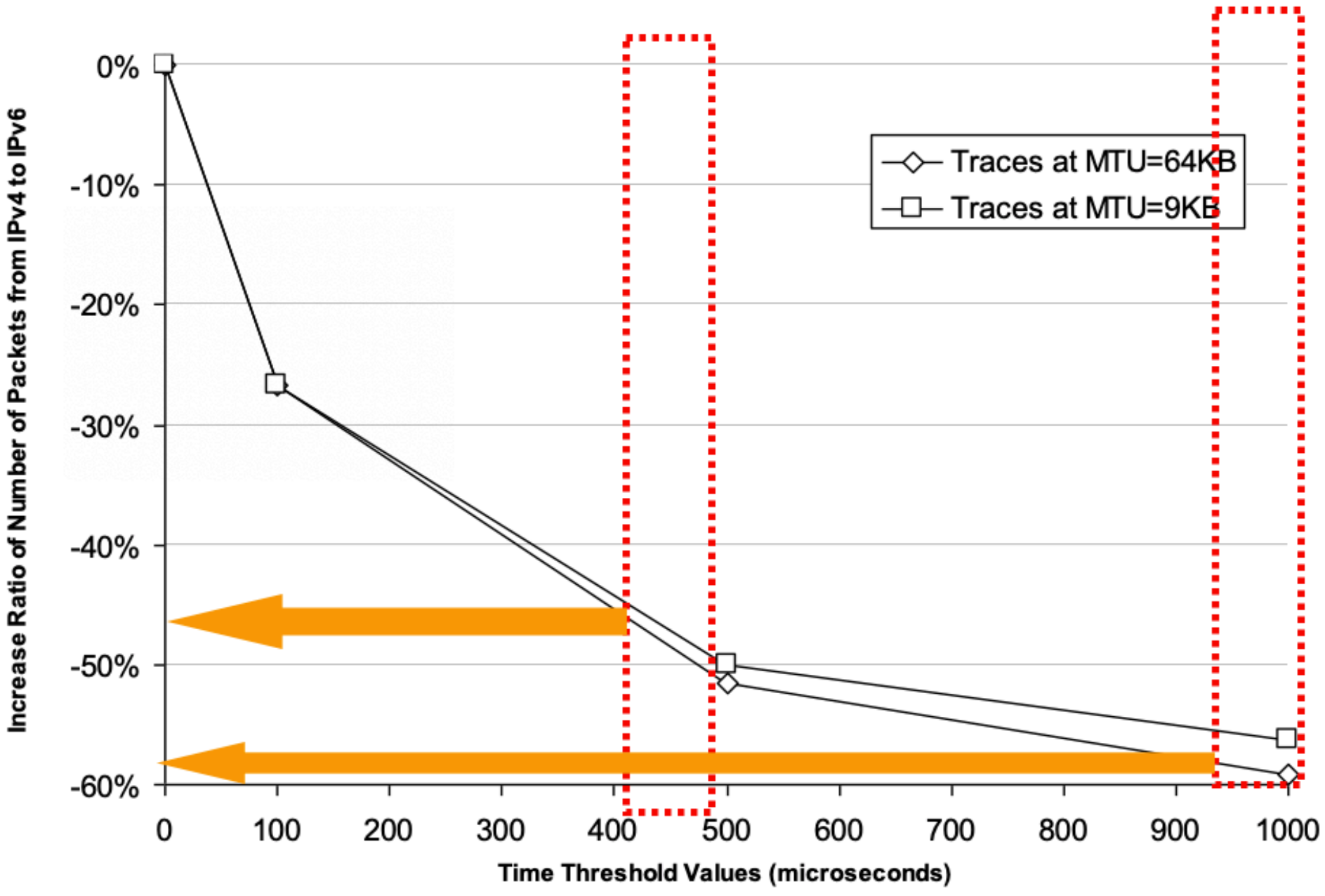

*Figure 10 – Different time aggregation thresholds and resulting increase / decrease in the number of generated packets, when converting IPv4 to IPv6, for a maximum packet size of 9 KB or 64KB, for the overall data traces (from [38]).*

## 4.5. Conclusions

Although the TCP/IP protocol stack has several limitations, this chapter was devoted to address only some of them, in particular the ones that result of the very small payload size of the standard Ethernet frame.

Changing all the machines that use Ethernet to a more efficient Data Link layer protocol seems something that will not occur in our lifetime, particularly because of the pervasiveness of the Ethernet-related formats all over the networking world, in particular, in the Internet.

Yet, and although this section is devoted to the conclusions, it has to be mentioned that in the early days of the new millennium, there have been several attempts to redesign the Internet. Notably, the first attempt was the "Clean Slate Internet" proposed at Stanford University. Oddly enough, the site webpage (https://cleanslate.stanford.edu) was reporting a `404 Error` and forwarding to the https://platformlab.stanford.edu/ website (last attempted access at the 26 of June of 2020). Nevertheless there are online reports at Stanford who mention this movement (*e.g.* [55]). Other initiatives such as *e.g.* the GENI (Global Environment for Network Innovation) [56], [57] or FEDERICA [58] (Federated E-infrastructure Dedicated to Research Innovating in Computing Architectures) and more recently, the Networld 2020 European Union project [59], also addressed the problem initially acknowledged by the Clean Slate, although with different approaches.

Having identified several issues and proposed some solutions, this chapter also presented research results that suggest that larger Layer 2 data units would result in a significantly lower number of packets being transmitted and routed in the network. Yet, more research needs to be done in order to fully assess the usefulness of such proposals.

(this page is intentionally left blank)

# 5. Keyed IPv6 protocol for connectionless segment transport

## 5.1. Introduction

The previous chapter was dedicated to address some of the limitation imposed onto the Internet Protocol because of the limitation of the size of the Layer 2 payload size. In this chapter a proposal to address the problem of connectionless communications at the Transport layer (Layer 4) is briefly presented, discussing an extension of this proposal to the Internet Protocol layer.

Layer 4 communications have been classified as Connection-oriented or Connectionless, as this layer is responsible for assuring the integrity of data transmission, or on the contrary, to not assure data transmission integrity at all. While at lower layers the data units are discarded if an error is detected, such as *e.g.* in the event of a frame collision at the transmission medium, or because the transmitted data has been corrupted by noise in the channel or malfunctioning equipment, the protocol stack layer that assures that the file received at the destination machine and application is an exact copy of the file at the source machine is the Transport layer. This is known as a Connection-oriented data transmission, and it mostly implemented over the Transmission Control Protocol (TCP) [5].

Yet, the Transport layer provides other types of protocols that can be used by applications that need to sacrifice the trustworthiness of the data transmission in benefit of data transmission speed, and this usually is the case for applications that provide services such as *e.g.* Voice over IP (VoIP) or video-conferencing services. Many of these protocols are industry proprietary, but eventually they fall under the category of the Connectionless data transmission protocols, the most prominent example being the User Datagram Protocol (UDP) [60].

One of the most distinctive differences between these two classes of protocols is that while TCP-like protocols rely on acknowledgments of successful data transmission to implement data transmission fidelity, the UDP-like protocols do not implement data transmission success metrics, and therefore, data is received by the destination machine only as successfully as the network was able to transport it, this meaning that the destination machine may receive packets that are out of sequence regarding their original order, and that the destination machine may miss packets without having the ability to identify any of these events.

Connectionless data transmissions have also a lower transmission overhead when compared to connection-oriented data transmissions, mostly because of two facts, the first being that the acknowledgment packets that connection oriented packets exchange do not contribute to the transmission of the application layer data but otherwise increase the data transmission overhead and delay, and also because the transport header of connectionless data transmissions is smaller than its connection-oriented counterpart, as connection-oriented headers need to include acknowledgment and sequence numbers (among other data) that allow the identification of out-of-sequence and missing packets.

The consequences of the usually termed "high overhead", "slow" TCP versus the "low overhead" and "light" UDP protocols are a well-studied subject and have given many contributions to network science, including *e.g.* many different "flavours" of TCP protocols, such as Reno, Vegas, among others [61]–[64]. But however well researched a subject is, there is always room for innovation, and this was what the concept of the Keyed User Datagram Protocol (KUDP) [64] proposes.

This chapter is dedicated to present the KUDP concept, an almost-reliable, zero added overhead regarding UDP, and its extension to the use of a similar *modus operandi* to IPv6, and the presentation of early performance assessment of these protocols regarding the ability to reorder packets that were received out-of-sequence, and to identify packets that were eventually lost during data transmission. For multimedia transmissions, this would eventually allow applications to apply data imputation techniques to minimize the effects of the lost packets, or, on other scenarios, to allow the applications to implement remediation strategies such as *e.g.* lowering resolution of the transmitted data.

### 5.2. The Keyed User Datagram Protocol

It is not the goal of this section to fully describe the concept and operation of KUDP, as this is best described in [64]. But a brief introduction has to be done to further extend the concept of Keyed IPv6 presented in section 5.3.

Figure 11 depicts a sequence of packets being transmitted from a source machine-application to a destination machine-application, using the UDP protocol. In this communication, the packets were generated with the same IP-Port source and destination pairs.

As the UDP header does not include segment nor acknowledgement numbers, it is impossible for the destination machine to identify and act upon network errors such as packets arriving out-of-sequence or packets that do not arrive at all.

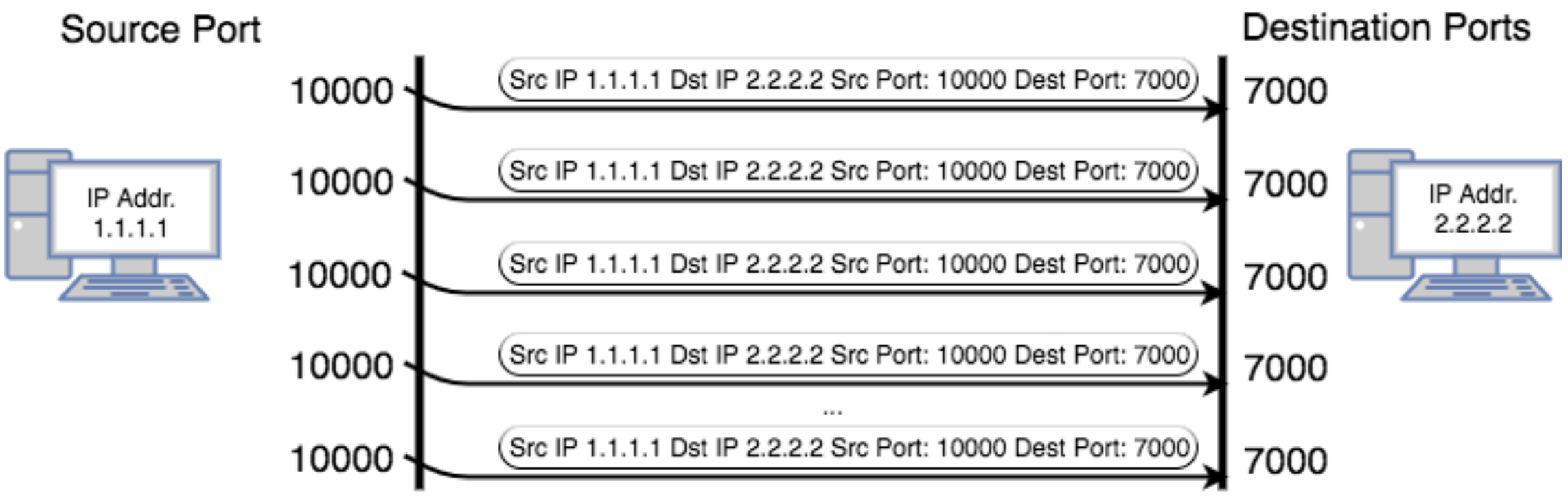

*Figure 11 – Scheme depicting the sending of an ordered set of several packets between two machines, using UDP (from [64]).*

We can now consider that because of the new protocol both the source and destination machines understand that the series of packet are not to be sent to the same port, but instead, they are to be sent sequentially and in a round-robin manner to a series of ports, *e.g.* ports 7000 to 7004, as shown in the example depicted in Figure 12. These destination ports are said to be the key of the protocol, in an analogy to a key that needs to address all the pins in the lock in the correct manner to allow the lock to open. The range of ports that compose the key of the protocol can be defined in the destination machine, or in the source machine, or in both, in what [64] describes as source-keyed, destination-keyed (example shown in Figure 12) or source-destination-keyed transmission. The way in which the pre-arrangement of the ports in the transmission is made is also discussed in [64], here considering three scenarios, depending on the which side (source, destination) the key is defined. The first one, as the vast majority of applications do, the port numbers are hard-coded into the applications' server and client. The second one would imply an inference algorithm to identify the key. The third one would add overhead to the transmission as it would require the use of a Key Definition Protocol.

In KUDP, if a packet is lost in the transmission, as the example in Figure 13 depicts, the destination machine, will notice that the received packets have ports that do not agree to the known key. The KUDP also allows the destination machine to easily reorder the packets that were received out-of-sequence.

Among the features of the KUDP, it is also important to point that when the key is applied at the source, *i.e.*, the transmission is sent from a series of ports from the source machine to a single destination port at the destination machine, the application at the destination machine can still receive and de-encapsulate the data even if it’s not implementing KUDP, *i.e.,* sKUDP is fully compatible with UDP.

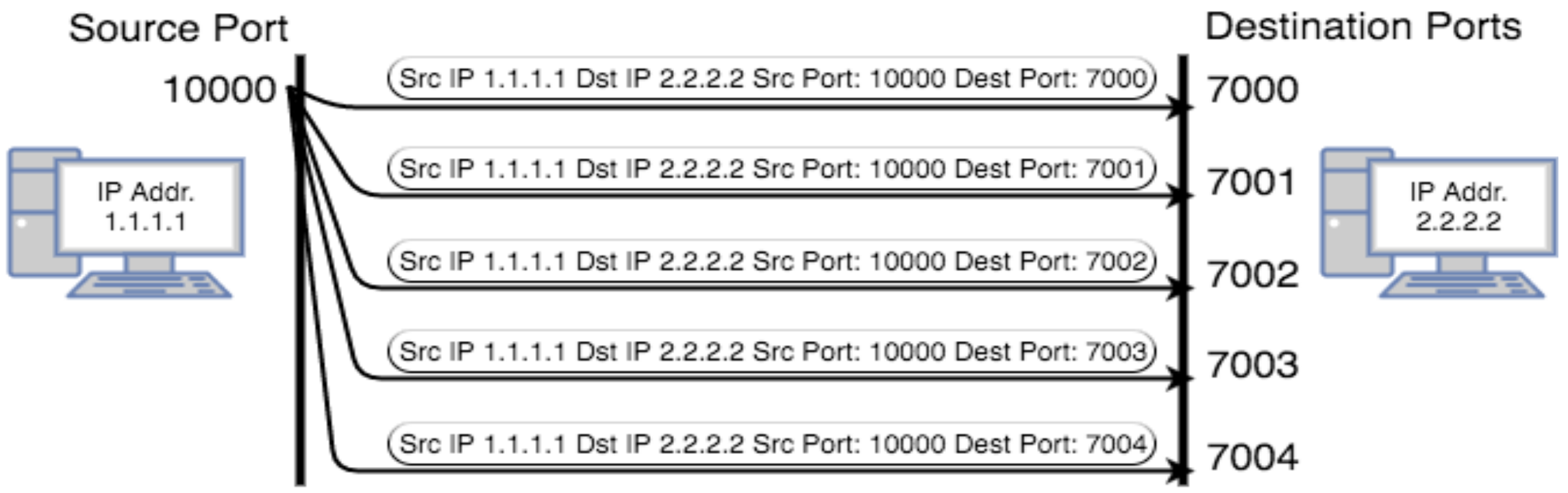

*Figure 12 – Scheme depicting the sending of an ordered set of 5 packets between two machines, using Keyed UDP (from [64]).*

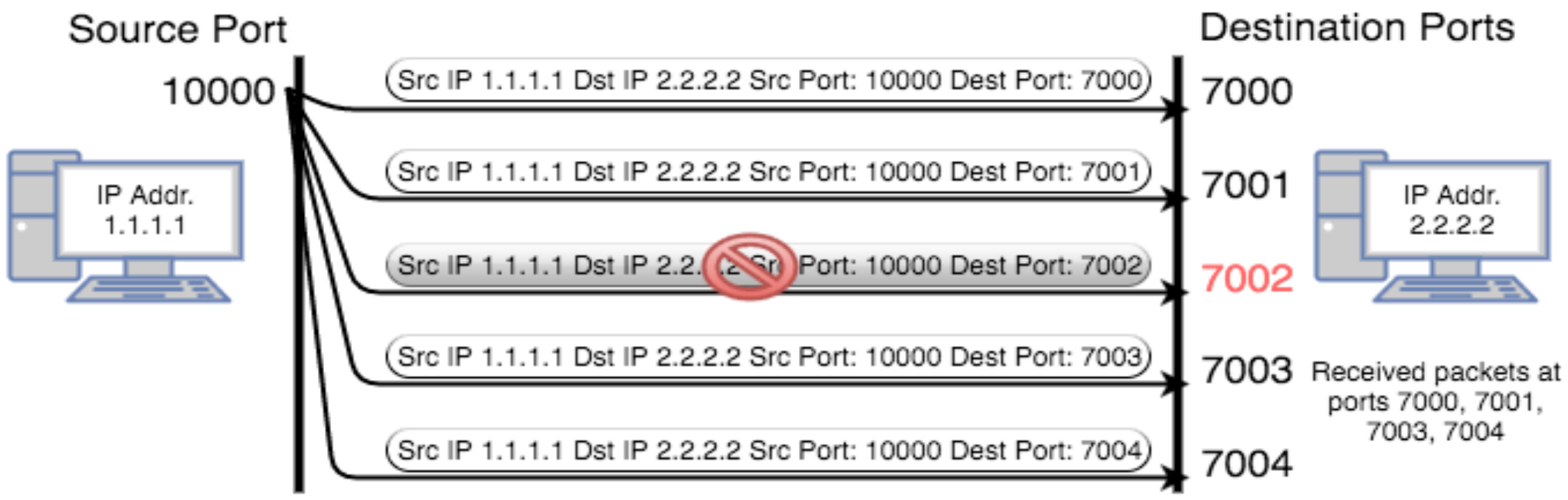

*Figure 13 – Basic loss sample scheme, depicting a transmission with a lost packet (from [64]).*

Some of the expected problems authors in [64] point, are related to the use of UDP through Network Address Translation – Port Translation (NAT-PT) machines. First, the use of additional ports from the source or to the destination machine will likely increase the number of the translation table entries in a NAT-PT machine. However, the impact this may have in the performance of the NAT-PT machine is unclear given the expectable size of the key and the number of ports that a single machine can use at any given moment in time. As an example, at the moment this document is being written, the author’s machine has registered the use of 601 ports.

Another problem regarding the presence of NAT-PT machines that may be encountered in the middle of a KUDP communication is a consequence of the manner in which NAT-PT machines work. These machines create a pair composed by a public IP address + port for each outgoing communication, based on the pair source-private IP + source-port. This means that in the case of source-keyed communications, the destination machine will very probably receive a set of source port numbers that does not correspond at all to initial port numbers defined by the source machine. If the destination machine is using a port inference algorithm to identify the source key, and the NAT-PT machine is using a fairly time-stable homomorphic algorithm, then there should be no difficulties. For destination-keyed transmissions the NAT-PT machine will not be a problem, as these machines do not alter the destination IP nor Port addresses. Again, the impact of NAT-PT machines in the source-keyed communications needs to be further researched.

### 5.3. The Keyed IPv6 Protocol

Authors of [64] also suggested the KUDP approach could be applied by defining IPv6 keys instead of port keys. Although the section of the paper where this is suggested is just two paragraphs long, authors state that all the considerations regarding KUDP could be applied to Keyed-IPv6, and added that eventually for long keys this would mean an overload to the ARP tables at the source and destination machines, particularly for communications within the same network [64]. Keyed-IPv6 relies on the concept that a single interface can be configured to respond to several IPv6 addresses.

In a Keyed-IPv6 communication, it is also possible to have source-, destination- and source-destination-keyed-IPv6, termed sK-IPv6, dK-IPv6 and sdK-IPv6, in a similar manner to the naming convention adopted in [64]. For sK-IPv6, a range of IPv6 addresses is used to send data to a single IPv6 address; for dK-IPv6, a single IPv6 address is used to send data to a range of IPv6 addresses; for sdK-IPv6, data is sent from a range of IPv6 from the source machine to a range of IPv6 addresses at the destination machine.

As the IPv6 addresses will belong to a well-defined network, *i.e.*, they belong to a range of addresses that was assigned to that interface, it not expected that this type of communication has impact in the routing tables of intermediate routers, unless the routers implement a very unlikely scenario of static routing with `/128` addresses. Moreover, if the used addresses at both machines are Global Unicast, the transmission will be transparent from beginning to end, and therefore will not be affected by address translation devices that it may encounter in the middle.

The Stream Reconstruction Algorithm described in [64] can be applied to Keyed-IPv6 without modification. The algorithm fully described in [64] consists in the following steps:

i. for a key that is *n* ports or IP addresses long, a buffer of *n-1* received packets is built; this will be queue number *1;*

ii. the packets in the buffer are sorted having the port number or the IP address as sort key; if a given value for the key is missing, the position if filled with a *-f-* value, for *failure to receive*;

iii. if present, the first packet is elected as the first packet of the stream; if absent, the first packet is declared as missing;

iv. define position to be set to 2, *j=2;*

v. start iteration on *j*

   a. the first position of the current queue *j-1* is removed and the current queue is duplicated; this is now queue *j*;
   b. a new packet is added to queue *j* and placed in its position, eventually overriding an *-f-* position;
   c. the packet for the position *j* is elected from all the candidates in all queues, from queue *j* to queue *j-n*;
   d. increment *j*;
   e. return no step *v.* until no more packets are received (timeout) and all buffers are processed.

Figure 14 shows the first nine iterations for the Stream Reconstruction Algorithm. The first column contains the transmitted packets, and the second column contains the received packets. Packets are identified as *1a, 2a, 3a, ..., 6a, 1b, 2b, ...*, and so on, but this is only to make it clearer to analyze the data. In this identification, the numbers 1, 2, 3, … represent the first, second, third, … address the packets were received by, as *a, b, c, d, ...* represent the round they were transmitted, *e.g.* packet *5c* was the third packet (the packet sent in the third round) sent to the 5$^{th}$ address. In Figure 14, packet tags in orange background stand for lost packets, and tags in yellow background stand for packets that were received out-of-sequence.

In Figure 14 it can also be seen that the receive buffers (marked in gray background in each iteration) are always *n-1* packets in length. Yet, and because the sorted buffer needs to place the packets in its expected position, this buffer may be bigger than *n-1.* For example, in the 4$^{th}$ iteration shown in Figure 14, the receive buffer is 5 packets in size, but the sorting buffer has 9 positions, to be able to accommodate the empty positions for the packets that did not arrive yet.

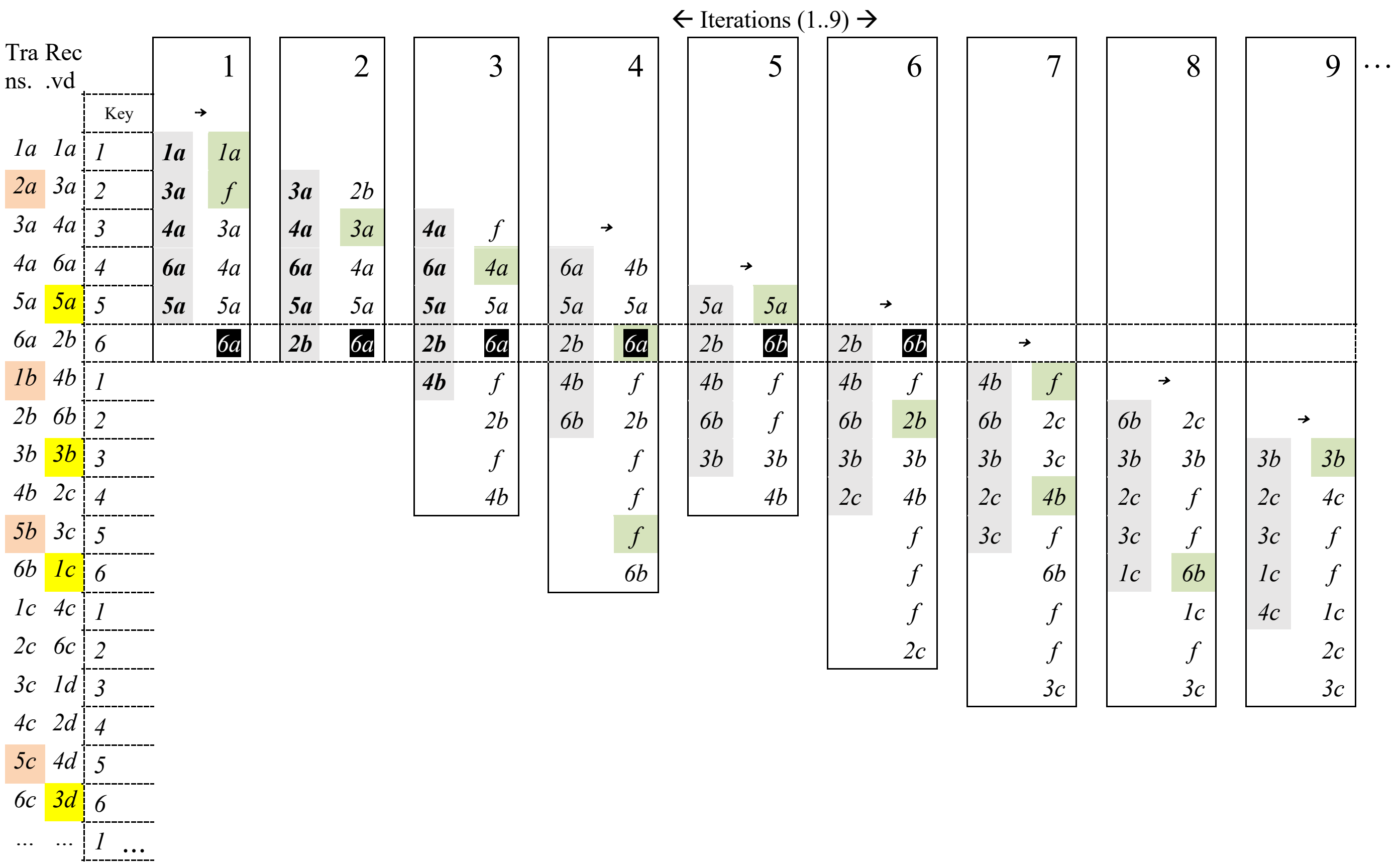

*Figure 14 – First nine iterations and corresponding sorting sets. At the left of each iteration, the received set of packets (gray background), at the right of each iteration, the sorted set of received packets, showing eventual missing packets. In black background, the candidates for the election for packet number 6. In green background, the elected packet (adapted from [64]).*

Also, as an example, it can be seen that for iteration 6, the sorting buffer is ordered as {*6b, f, 2b, 3b, 4b, f, f, f, 2c*}. This buffer will be first used to elect the packet that will occupy the 6th position on the final stream. The election of the 6th packet is marked in the figure with a dotted horizontal frame. For each election, there are *n-1* sorting buffers to be considered. The candidates for this election are packets in the set {*6a, 6a, 6a, 6a, 6b, 6b*}, therefore, candidate *6a* wins.

Authors in [64] also warn on the possibility of a hidden packet, *i.e.*, a packet that despite being received is never placed in its position in the queue, because that position was already occupied. The election rules are as follows:

i. the candidate with most occurrences wins;
ii. in case of a tie, the first occurrence for that candidate wins;
iii. in case of a hidden packet, fit the packet in the first corresponding free slot;
iv. finally, once elected, a candidate cannot be eligible again, *i.e.*, no duplicates are allowed.

In the example in Figure 14, the first nine received packets were {*1a, 3a, 4a, 6a, 5a, 2b, 4b, 6b, 3b*}. Applying the Stream Reconstruction Algorithm described previously, the output of the algorithm after the 9$^{th}$ iteration is {*1a, f, 3a, 4a, 5a, 6a, f, 2b, 3b*}, *i.e.*, the algorithm was able to identify that the packets *2* and *1* from the first and the second round respectively (here marked as *a* and *b*) were missing, and was able to correctly reorder packets that arrived out-of-sequence, *i.e.* packets *6a* and *5a*, and packets *6b* and *3b.*

It should be noted that Keyed-IPv6 and KUDP do not propose any manner to recover the lost packets nor to inform the source machine of the efficiency of the data transmission, as the main focus of this concept was to allow the destination machine to have as much information as possible on the success of the transmission with a zero added overhead when compared to standard UDP.

But, not maintaining the full compatibility with UDP, in the case of a proprietary protocol, it is possible to achieve the identification of lost packets and the reordering of packets received out-of-sequence by, for example, adding sequence or control data that could trail or lead the application data, still keep the overhead to a minimum.

### 5.4. Results

The efficiency for the KUDP and Keyed-IPv6 algorithms was tested through simulation [65]. When [64] was published, several topics were pointed as future research, being one of them the optimum size for a key, as it becomes clear that longer keys allow for a better identification of errors, but also can impose a higher load on operating systems and on the Stream Reconstruction Algorithm. Another aspect pointed as future research was the size of the buffer as a function of the size of the key for the Stream Reconstruction Algorithm. In the previous section it was defined that the buffer should hold *n-1* packets, being *n* the size of the key. This value is described in [64] as the result of a basic set of preliminary experiments.

Research in [65] has addressed these three questions, in particular, regarding the efficiency of the Stream Reconstruction Algorithm:

1) for key sizes of 5, 10, 15, 20, 50, 100 and 200 positions;
2) for arrivals out-of-sequence or for missing packets; and
3) for the size of the iteration buffer as function of the size of the key.

After the initial research in [64], several buffer sizes were experimented with and it was found that keeping the buffer as half the size of the key (rounded by excess) provided results that were satisfactory. Also, in the performed simulations, it was decided to separate the two events, *i.e.*, the simulations tested either the arrival of packets out-of-sequence or the event of missing packets, but not the existence of both events.

*Table 12 – Efficiency of the Stream Reconstruction Algorithm for ratios of packets received out-of-sequence ratios versus key length (from [65]).*

| | 10% | 35% | 50% | 60% | 70% |
|---|---|---|---|---|---|
| **n=5** | 95.95% | 81.85% | 76.90% | 75.50% | 75.10% |
| **n=10** | 100.00% | 99.12% | 95.92% | 92.37% | 90.07% |
| **n=15** | 100.00% | 99.83% | 98.76% | 97.61% | 94.24% |
| **n=20** | 100.00% | 99.92% | 99.63% | 99.06% | 97.53% |
| **n=50** | 100.00% | 100.00% | 100.00% | 99.99% | 99.97% |
| **n=100** | 100.00% | 100.00% | 100.00% | 100.00% | 99.99% |
| **n=200** | 100.00% | 100.00% | 100.00% | 100.00% | 100.00% |

*Table 13 – Efficiency of the Stream Reconstruction Algorithm for lost packets ratios versus key length (from [65]).*

| | 10% | 15% | 20% | 25% |
|---|---|---|---|---|
| **n=5** | 99.75% | 99.10% | 97.50% | 95.40% |
| **n=10** | 99.88% | 99.50% | 97.82% | 96.00% |
| **n=15** | 99.97% | 99.90% | 99.25% | 96.30% |
| **n=20** | 100.00% | 99.98% | 99.25% | 97.15% |
| **n=50** | 100.00% | 100.00% | 99.70% | 97.57% |
| **n=100** | 100.00% | 100.00% | 99.93% | 98.38% |
| **n=200** | 100.00% | 100.00% | 100.00% | 99.10% |

These were the simulation parameters used:

a) for each simulation run, the number of streams of packets was defined as 4 times the size of the key. This is enough as there is never a situation in which packets from all the 4 rounds of present at the election time;

b) Packet loss thresholds were defined from ranging 1% up to an unrealistic 25%. For each generated packet, a random number is associated, therefore allowing for the decision if that packet will be lost or not, and therefore, the exact packet loss ratios for each simulation run do not coincide exactly with the defined threshold;

c) Out-of-sequence arrival rates were defined as ranging from 1% up to 70%. In a load balancing routing situation with two equal cost paths, if one of the routers systematically delays its routing packets, it is expectable to have a 50% ratio for packets that arrive out-of-sequence, and therefore, the extension of the simulation threshold up to 70%;
d) Each simulation was run 100 times, and the presented results show the average for all the runs.

Tables 12 and 13 show some of the simulation results for the beforementioned scenarios. It can be seen that for a key length of 20, the Stream Reconstruction Algorithm is able to identify and correct all of the 10% of out-of-sequence packet received, but for a key length of 50, the algorithm is able to place in order all of the 50% of packets that arrive out-of-sequence (and 99.99% of the packets if 60% of the packets arrive out-of-sequence).

The algorithm is said to be efficient when it detects a lost packed as missing and when it is able to place in-sequence packets that were received out-of-sequence.

### 5.5. Conclusions

This chapter presented research concerning the extension of the KUDP to IPv6. The KUDP protocol relies on the use of a sequence of ports to allow the receiving machine to perform two tasks that are unreachable for UDP-like protocols: the identification of packets that were lost, and the reordering of packets that arrived out-of-sequence. Almost-reliable protocols such as the KUDP or the K-IPv6 allow upper network layers to assess the quality of the data packet transmission, as early simulation results show.

The next steps, already started, consist in producing a real-time communication tool based in KUDP and in K-IPv6 and test it in a real wide area network scenario. This will allow to further build on the KUDP concept to allow *e.g.* the exchange of low overhead messages that allow the exchange of information between the source and the destination terminal, or to allow for data imputation in multimedia streams.

# 6. Conclusions and future work

## 6.1. Introduction

This document complies to the required by point 2. c) of Article 8 of the Decree Law 239/2007 published in the *Diário da República, 1ª Série, Nº 116, de 19 de Junho de 2007*, and contains the Detailed Summary of the Seminar addressing the subject of "The Internet Protocol, past, some current limitations and a glimpse of a possible future".

The previous chapters have provided a discussion of the history of the development of the Internet Protocol, since the early days of the Network Control Protocol up until its version 6, of some of the limitations of the Internet Protocol in a broader sense, since the IP is not limited in itself, but it has been limited by both the upper and lower layers, and finally, it presents research pointing some possible solutions to specific aspects of these limitations, presenting a proposal for future IP scenarios. It also presents a new method to perform IP address assignment in subnetworks.

## 6.2. Future work

The detailed summary of the seminar presents some issues that constitute future work. The concept of "Everything over IP" and "IP over Everything", which jointly aimed to simplify the protocol stack, promote interoperability among heterogeneous connected devices, and eventually, reduce some of the problems particularly imposed by old Layer 2 technologies, set the tone for the discussion of this Future work.

The limitations the layer 2 technologies impose onto the upper layers is a well-studied field, and as this seminar is focused on Layer 3 and IP, these fall out of the scope of this document. Yet, it needs to be noted that, with IPv6, the most obvious limitations of the IP protocol were solved, in particular the most prominent being the small pool of addresses of IPv4, but also the fact that IPv4 had no built-in security features, which are now part of the IPv6 standard.

Layer 2.5 technologies such as the Multi Protocol Label Switching (MPLS) also fall out of the scope of this seminar, and its use is currently implemented to achieve, among other things, faster packet transmission in the transport / core networks, easier scalability of virtual circuits and easier management by means of the Software Defined Network (SDN) architectures.

So, in fact, IPv6 seems to conceal no particular limitation in itself. It has a virtually inexhaustible range of addresses, it integrates security features, and it is implemented in all major operating systems. Being all good news, there are still open fields of research, as Chapter 5 discusses.

Current ongoing research, hosted at the *Instituto de Telecomunicações / Universidade da Beira Interior* and lead by the author, aims to produce a proof-of-concept of the new K-IPv6 protocol. This research work is currently assigned to two master students. The first student has built a simulator that after implementing the Stream Reconstruction Algorithm, allowed the simulation of a data transmission in KUDP (and K-IPv6 by extension), and whose results were presented in Section 5.4. The goal of this research is now focused on finetuning the Stream Reconstruction Algorithm, as to improve its efficiency with smaller keys. Additionally, the research will also address the issue of the latency imposed onto the packets because of the election process in the algorithm. Other issue that will also be addressed is the changes the Stream Reconstruction Algorithm will cause to jitter. The initial results of the efficiency of the KUDP or K-IPv6 are currently submitted to an IEEE networking conference.

Despite the good simulation results, the real efficiency of the KUDP or K-IPv6 needs to be assessed by a video or audio real-time data transmission tool tested in a real environment. In particular, we want to be able to assess the resiliency of the Stream Reconstruction Algorithm of this type of almost-reliable protocols, and address topics such as communication start and ending, the feasibility of asking for the retransmission of a particular packet, and to explore the possibility of using this protocol as an almost reliable TCP. This is the task of the second master student, whose goal is to develop a proof-of-concept real time communication application. This line of research has recently begun, and in part depends on the conclusion of the previous research on the finetuning of the Stream Reconstruction Algorithm.

Other topics of research regarding the kUDP and the K-IPv6 are its resilience to intermediate machines such as NAT-PT or Firewalls. This line of research will be conducted after the prototype for the real time communication tool has been implemented and validated.

A final topic of research in this line is the feasibility of joint use of kUDP and K-IPv6 to convey masked packet data, as such a sequence of source-destination IP addresses and port numbers would render almost impossible the identification and interruption of a data transmission. Yet, this research can only start when the application prototype is operational.

### 6.3. Closing remarks

This document has presented the detailed summary of the seminar intitled “The Internet Protocol, past, some current limitations and a glimpse of a possible future”. This was done in a manner that was not risk free. Despite the attempts to contain functionalities as much as possible in each of the TCP/IP layers, it becomes clear that in the end, the limitations of the physical channel, *i.e.*, the limitations resulting from the frailty of the transmission of each bit travelling from the transmitting to the receiving machine, affect the way the upper layers perform their tasks.

As an allegory to many aspects of life itself, often what is limited by the design of the foundations, spreads all the way to the top, defining what otherwise well architectured solutions can deliver. Taking into consideration the pervasiveness of the Ethernet hardware, it is very hard to envision a networking scenario that is not populated by tiny-grams, and therefore, it is expected that we will need to relying on complex strategies and architectures to simplify core routing as our machines will continue to generate thousands of packets just because we need send a picture as an attachment in an email.

(this page is intentionally left blank)

# References

[1] "A Brief History of the Telephone Number | GetVoIP." https://getvoip.com/blog/2012/11/15/a-brief-history-of-the-telephone-number/ (accessed Oct. 13, 2019).
[2] "§ 91. A short history of telephone numbers." https://www.artlebedev.com/mandership/91/ (accessed Oct. 13, 2019).
[3] "Brief History of the Internet," *Internet Society*. https://www.internetsociety.org/internet/history-internet/brief-history-internet/ (accessed Oct. 10, 2019).
[4] C. Vint and R. Kahn, "A protocol for packet network interconnection," *IEEE Trans. Commun.*, 1974.
[5] Information Sciences Institute, *Transmission Control Protocol*, vol. RFC 793. 1981.
[6] "Version Numbers." https://www.iana.org/assignments/version-numbers/version-numbers.xhtml (accessed Oct. 10, 2019).
[7] G. Swallow, L. Andersson, and S. Bryant, "Avoiding Equal Cost Multipath Treatment in MPLS Networks." https://tools.ietf.org/html/rfc4928 (accessed Oct. 10, 2019).
[8] R. Jonathan Postel, *Internet Protocol Specification*, vol. IEN 28. 1978.
[9] Jon Postel, "Comments on Internet Protocol and TCP." https://www.rfc-editor.org/ien/ien2.txt (accessed Oct. 10, 2019).
[10] S. C. Herring, "Should You Be Capitalizing the Word 'Internet'?," *Wired*, Oct. 19, 2015.
[11] J. Postel, "RFC 791: Internet protocol," 1981.
[12] James W. Forgie, "Internet Stream Protocol." https://www.rfc-editor.org/ien/ien119.txt (accessed Oct. 10, 2019).
[13] C. Topolcic, "Experimental Internet Stream Protocol: Version 2 (ST-II)." https://tools.ietf.org/html/rfc1190 (accessed Oct. 11, 2019).
[14] "Update 2016: What Happened to IPv5?," *Colocation America*, Aug. 21, 2015. https://www.colocationamerica.com/blog/ipv4-ipv6-what-happened-to-ipv5.htm (accessed Oct. 11, 2019).
[15] J. Postel and J. Reynolds, "Assigned Numbers (RFC 1700)." https://tools.ietf.org/html/rfc1700 (accessed Oct. 11, 2019).
[16] J. Postel and J. K. Reynolds, "Assigned numbers (RFC 870)." https://tools.ietf.org/html/rfc870 (accessed Oct. 11, 2019).
[17] G. J. de Groot, Y. Rekhter, D. Karrenberg, and E. Lear, "Address Allocation for Private Internets." https://tools.ietf.org/html/rfc1918 (accessed Oct. 12, 2019).
[18] Y. Rekhter, D. Karrenberg, G. de Groot, and B. Moskowitz, "Address Allocation for Private Internets." https://tools.ietf.org/html/rfc1597 (accessed Oct. 12, 2019).
[19] D. Clark, R. Braden, L. Chapin, R. Hobby, and V. Cerf, "Towards the Future Internet Architecture." https://tools.ietf.org/html/rfc1287 (accessed Oct. 12, 2019).
[20] E. Lear, E. Fair, T. Kessler, and D. Crocker, "Network 10 Considered Harmful (Some Practices Shouldn't be Codified)." https://tools.ietf.org/html/rfc1627 (accessed Oct. 12, 2019).
[21] J. Yu, K. Varadhan, T. Li, and V. Fuller, "Classless Inter-Domain Routing (CIDR): an Address Assignment and Aggregation Strategy." https://tools.ietf.org/html/rfc1519#section-1 (accessed Oct. 13, 2019).

[22] J. Yu, K. Varadhan, T. Li, and V. Fuller, "Supernetting: an Address Assignment and Aggregation Strategy." https://tools.ietf.org/html/rfc1338 (accessed Oct. 13, 2019).
[23] V. Fuller and T. Li, "Classless Inter-domain Routing (CIDR): The Internet Address Assignment and Aggregation Plan." https://tools.ietf.org/html/rfc4632#page-5 (accessed Oct. 13, 2019).
[24] R. Hinden and S. Deering, "Internet Protocol, Version 6 (IPv6) Specification (RFC 1883)." https://tools.ietf.org/html/rfc1883#page-33 (accessed Oct. 13, 2019).
[25] R. Hinden and S Deering, "IP Version 6 Addressing Architecture." https://tools.ietf.org/html/rfc2373#page-5 (accessed Oct. 13, 2019).
[26] "RFC 8200 - IPv6 has been standardized," *Internet Society*, Jul. 17, 2017. https://www.internetsociety.org/blog/2017/07/rfc-8200-ipv6-has-been-standardized/ (accessed Oct. 16, 2019).
[27] "World Population Clock: 7.7 Billion People (2019) - Worldometers." https://www.worldometers.info/world-population/ (accessed Oct. 20, 2019).
[28] E. Bianconi *et al.*, "An estimation of the number of cells in the human body," *Ann. Hum. Biol.*, vol. 40, no. 6, pp. 463–471, Dec. 2013, doi: 10.3109/03014460.2013.807878.
[29] R. M. Hinden and S. E. Deering, "IP Version 6 Addressing Architecture (RFC 4191)." https://tools.ietf.org/html/rfc4291 (accessed Mar. 07, 2020).
[30] R. M. Hinden and B. Haberman, "Unique local IPv6 unicast addresses," 2005.
[31] R. M. Hinden and S. E. Deering, "Internet Protocol Version 6 (IPv6) Addressing Architecture." https://tools.ietf.org/html/rfc3513 (accessed Mar. 07, 2020).
[32] P. Smith, G. Huston, and A. Lord, "IPv6 Address Prefix Reserved for Documentation." https://tools.ietf.org/html/rfc3849 (accessed Mar. 07, 2020).
[33] C. Huitema and B. Carpenter, "Deprecating Site Local Addresses." https://tools.ietf.org/html/rfc3879 (accessed Mar. 07, 2020).
[34] B. D. Zill, T. Jinmei, S. E. Deering, E. Nordmark, and B. Haberman, "IPv6 Scoped Address Architecture." https://tools.ietf.org/html/rfc4007 (accessed Mar. 07, 2020).
[35] J. Arkko and F. Baker, "Guidelines for Using IPv6 Transition Mechanisms during IPv6 Deployment." https://tools.ietf.org/html/rfc6180 (accessed Mar. 08, 2020).
[36] D. G. Waddington and F. Chang, "Realizing the transition to IPv6," *IEEE Commun. Mag.*, pp. 139–144, Jun. 2002.
[37] R. E. Gilligan and E. Nordmark, "Basic Transition Mechanisms for IPv6 Hosts and Routers." https://tools.ietf.org/html/rfc4213 (accessed Mar. 08, 2020).
[38] N. M. Garcia, M. M. Freire, and P. P. Monteiro, "The Ethernet Frame Payload Size and Its Effect on IPv4 and IPv6 Traffic," 2008, pp. 1–5.
[39] "IP Addressing and Subnetting for New Users," *Cisco*. https://www.cisco.com/c/en/us/support/docs/ip/routing-information-protocol-rip/13788-3.html (accessed Mar. 08, 2020).
[40] T. T. Pummill <trop@alantec.com>, "Variable Length Subnet Table For IPv4." https://tools.ietf.org/html/rfc1878 (accessed Mar. 08, 2020).
[41] J. Postel and J. C. Mogul, "Internet Standard Subnetting Procedure." https://tools.ietf.org/html/rfc950 (accessed Mar. 08, 2020).
[42] Y. Li, D. Li, W. Cui, and R. Zhang, "Research based on OSI model," in *2011 IEEE 3rd International Conference on Communication Software and Networks*, 2011, pp. 554–557.
[43] "The Ethernet - A Local Area Network, Data Link Layer and Physical Layer Specification," p. 120.
[44] "2 Ethernet — An Introduction to Computer Networks." http://intronetworks.cs.luc.edu/1/html/ethernet.html (accessed Jun. 21, 2020).
[45] "The History and Future of Internet Traffic," *Cisco Blogs*, Aug. 28, 2015. https://blogs.cisco.com/sp/the-history-and-future-of-internet-traffic (accessed Jun. 20, 2020).

[46] Q. Wang *et al.*, "Multimedia IoT systems and applications," in *2017 Global Internet of Things Summit (GIoTS)*, 2017, pp. 1–6.
[47] S. Bj, "Can OTN be replaced by Ethernet? A network level comparison of OTN and Ethernet with a 5G perspective," in *2018 International Conference on Optical Network Design and Modeling (ONDM)*, 2018, pp. 220–225.
[48] S. R. Amstutz, "Burst Switching - An Introduction," *IEEE Communications Magazine*, pp. 36–42, Nov. 1983.
[49] M. Freire, N. Garcia, M. Hajduczenia, P. Monteiro, and H. Silva, "Method and machine for aggregating a plurality of data packets into a unified transport data packet," EP1848172 (A1).
[50] N. M. G. dos Santos, "Architectures and Algorithms for IPv4/IPv6-Compliant Optical Burst Switching Networks," Universidade da Beira Interior, 2008.
[51] D. Borman, S. Deering, and R. Hinden, *IPv6 Jumbograms*, vol. RFC 2675. 1999.
[52] N. M. Garcia, P. Monteiro, and M. Freire, "Assessment of burst assembly algorithms using real ipv4 data traces," in *The 2nd International Conference on Distributed Frameworks for Multimedia Applications*, 2006, pp. 1–6.
[53] N. M. Garcia, P. P. Monteiro, and M. M. Freire, "Burst assembly with real IPv4 data–performance assessement of three assembly algorithms," in *International Conference on Next Generation Wired/Wireless Networking*, 2006, pp. 223–234.
[54] Ethernet Alliance, "Ethernet Jumbo Frames," 2011. http://www.ethernetalliance.org/wp-content/uploads/2011/10/EA-Ethernet-Jumbo-Frames-v0-1.pdf (accessed Apr. 24, 2013).
[55] "Clean Slate Internet." Accessed: Jun. 27, 2020. [Online]. Available: https://forum.stanford.edu/events/2008/2008slides/Cleanslate/POMI-workshop-opening-Guru.pdf.
[56] R. McGeer, M. Berman, C. Elliott, and R. Ricci, *The GENI book*. Springer, 2016.
[57] "GENI." https://www.geni.net/ (accessed Jun. 27, 2020).
[58] "Federated E-infrastructure Dedicated to European Researchers Innovating in Computing network Architectures | FEDERICA Project | FP7," *CORDIS | European Commission*. https://cordis.europa.eu/project/id/213107 (accessed Jun. 27, 2020).
[59] "NetWorld2020 ETP – European Technology Platform." https://www.networld2020.eu/ (accessed Jun. 27, 2020).
[60] J. Postel, "User datagram protocol," 1980.
[61] L. A. Grieco and S. Mascolo, "Performance evaluation and comparison of Westwood+, New Reno, and Vegas TCP congestion control," *ACM SIGCOMM Comput. Commun. Rev.*, vol. 34, no. 2, pp. 25–38, 2004.
[62] J. Mo, R. J. La, V. Anantharam, and J. Walrand, "Analysis and comparison of TCP Reno and Vegas," in *IEEE INFOCOM'99. Conference on Computer Communications. Proceedings. Eighteenth Annual Joint Conference of the IEEE Computer and Communications Societies. The Future is Now (Cat. No. 99CH36320)*, 1999, vol. 3, pp. 1556–1563.
[63] A. Detti and M. Listanti, "Impact of Segments Aggregation on TCP Reno Flows in Optical Burst Switching Networks," Jun. 2002.
[64] N. M. Garcia, F. Gil, B. Matos, C. Yahaya, N. Pombo, and R. I. Goleva, "Keyed User Datagram Protocol: Concepts and Operation of an Almost Reliable Connectionless Transport Protocol," *IEEE Access*, vol. 7, pp. 18951–18963, 2019.
[65] F. Machado Gil, N. M. Garcia, B. Matos, N. Pombo, R. Goleva, and C. Dobre, "Identifying Packet Loss and Reordering Packets in Keyed UDP Transmissions," *arXiv*, p. arXiv: 2007.00127, 2020.